\documentclass[twocolumn,times,trackchanges]{aastex63} 
\usepackage{longtable}
\usepackage{booktabs}

\newcommand{\Mni}{$M_{\rm Ni}$}

\newcommand{\kms}{km~s$^{-1}$}
\newcommand{\ergs}{erg s$^{-1}$}

\newcommand{\NaI}{Na~{\sc i}}

\newcommand{\MgII}{Mg~{\sc ii}}

\newcommand{\SiII}{Si~{\sc ii}}

\newcommand{\SII}{S~{\sc ii}}
\newcommand{\CaII}{Ca~{\sc ii}}

\newcommand{\FeII}{Fe~{\sc ii}}
\newcommand{\FeIII}{Fe~{\sc iii}}

\newcommand{\CoII}{Co~{\sc ii}}
\newcommand{\CoIII}{Co~{\sc iii}}
\newcommand{\NiII}{Ni~{\sc ii}}

\newcommand{\Nifs}{$^{56}$Ni}

\newcommand{\sBV}{$s_{BV}$}

\newcommand{\mb}{${\Delta}m_{15}(B)$}

\newcommand{\Angst}{\AA}

\newcommand{\RSiSi}{$R$(\SiII)}

\newcommand{\dmvalue}{$1.05 \pm 0.07$}
\newcommand{\sbvalue}{$1.19 \pm 0.03$}
\newcommand{\distmd}{$32.39 \pm 0.06$}
\newcommand{\mbmag}{$-19.32 \pm 0.13$}
\newcommand{\mB}{$13.07 \pm 0.11$}
\newcommand{\mV}{$12.91 \pm 0.07$}
\newcommand{\BVmax}{$0.16 \pm 0.13$}
\newcommand{\ebvalue}{$0.17 \pm 0.07$}
\newcommand{\tbmax}{$57,959.4 \pm 0.4$}
\newcommand{\tzero}{$57941.4 \pm 0.4$}
\newcommand{\tlcv}{$13.51 \pm 0.08$}
\newcommand{\tgama}{$38.59 \pm 2.28$}
\newcommand{\trise}{$18.0 \pm 0.4$}

\newcommand{\Lmax}{($1.32\pm\,0.13)\,\times 10^{43}$\,erg\,s$^{-1}$}
\newcommand{\MniValue}{$0.51 \pm 0.03\,M_{\odot}$}

\newcommand{\VSiII}{$15,000 \pm 150$\,\kms}
\newcommand{\VSiIIDot}{$120 \pm 10$\,km\,s$^{-1}$\,d$^{-1}$}
\newcommand{\RSiII}{$0.04 \pm 0.01$}

\begin{document}

\shorttitle{Observations of SN 2017fgc}
\shortauthors{Zeng et al.}

\title{SN 2017fgc: A Fast-Expanding Type Ia Supernova Exploded in Massive Shell Galaxy NGC 474}

\author{Xiangyun Zeng}  
\affil{Xinjiang Astronomical Observatory, Chinese Academy of Sciences, Urumqi, Xinjiang 830011, People's Republic of China}
\affil{School of Astronomy and Space Science, University of Chinese Academy of Sciences, Beijing 100049, People's Republic of China}

\author{Xiaofeng Wang}  
\email{wang\_xf@mail.tsinghua.edu.cn}
\affil{Physics Department and Tsinghua Center for Astrophysics (THCA), Tsinghua University, Beijing, 100084, People's Republic of China}
\affil{Beijing Planetarium, Beijing Academy of Science of Technology, Beijing 100044, People's Republic of China}

\author{Ali Esamdin}  
\email{aliyi@xao.ac.cn}
\affil{Xinjiang Astronomical Observatory, Chinese Academy of Sciences, Urumqi, Xinjiang 830011, People's Republic of China}

\author{Craig Pellegrino}  
\affil{Department of Physics, University of California, Santa Barbara, CA 93106-9530, USA}
\affil{Las Cumbres Observatory, 6740 Cortona Drive Suite 102, Goleta, CA 93117-5575, USA}

\author{Jamison Burke}  
\affil{Las Cumbres Observatory, 6740 Cortona Drive Suite 102, Goleta, CA 93117-5575, USA}
\affil{Department of Physics, University of California, Santa Barbara, CA 93106-9530, USA}

\author{Benjamin E. Stahl} 
\affil{Department of Astronomy, University of California, Berkeley, CA 94720-3411, USA}
\affil{Department of Physics, University of California, Berkeley, CA 94720-7300, USA}
\affil{Marc J. Staley Graduate Fellow}

\author{WeiKang Zheng} 
\affil{Department of Astronomy, University of California, Berkeley, CA 94720-3411, USA}

\author{Alexei V. Filippenko}  
\affil{Department of Astronomy, University of California, Berkeley, CA 94720-3411, USA}
\affil{Miller Senior Fellow, Miller Institute for Basic Research in Science, University of California, Berkeley, CA 94720, USA}

\author{D. Andrew Howell}  
\affil{Las Cumbres Observatory, 6740 Cortona Drive Suite 102, Goleta, CA 93117-5575, USA}
\affil{Department of Physics, University of California, Santa Barbara, CA 93106-9530, USA}

\author[0000-0003-4102-380X]{D. J. Sand}  
\affiliation{Steward Observatory, University of Arizona, 933 North Cherry Avenue, Rm. N204, Tucson, AZ 85721-0065, USA}

\author{Stefano Valenti}  
\affil{Department of Physics, University of California, Davis, CA 95616, USA}

\author{Jun Mo}   
\affil{Physics Department and Tsinghua Center for Astrophysics (THCA), Tsinghua University, Beijing, 100084, People's Republic of China}

\author{Gaobo Xi}    
\affil{Physics Department and Tsinghua Center for Astrophysics (THCA), Tsinghua University, Beijing, 100084, People's Republic of China}

\author{Jialian Liu}   
\affil{Physics Department and Tsinghua Center for Astrophysics (THCA), Tsinghua University, Beijing, 100084, People's Republic of China}

\author{Jujia Zhang}  
\affil{Yunnan Observatories (YNAO), Chinese Academy of Sciences, Kunming 650216, People's Republic of China}
\affil{Key Laboratory for the Structure and Evolution of Celestial Objects, Chinese Academy of Sciences, Kunming 650216, People's Republic of China}
\affil{Center for Astronomical Mega-Science, Chinese Academy of Sciences, 20A Datun Road, Chaoyang District, Beijing, 100012, People's Republic of China}

\author{Wenxiong Li} 
\affil{The School of Physics and Astronomy, Tel Aviv University, Tel Aviv 69978, Israel}

\author{Abdusamatjan Iskandar} 
\affil{Xinjiang Astronomical Observatory, Chinese Academy of Sciences, Urumqi, Xinjiang 830011, People's Republic of China}
\affil{School of Astronomy and Space Science, University of Chinese Academy of Sciences, Beijing 100049, People's Republic of China}

\author{Mengfan Zhang} 
\affil{Xinjiang Astronomical Observatory, Chinese Academy of Sciences, Urumqi, Xinjiang 830011, People's Republic of China}


\author{Han Lin}    
\affil{Physics Department and Tsinghua Center for Astrophysics (THCA), Tsinghua University, Beijing, 100084, People's Republic of China}

\author{Hanna Sai}   
\affil{Physics Department and Tsinghua Center for Astrophysics (THCA), Tsinghua University, Beijing, 100084, People's Republic of China}

\author{Danfeng Xiang}   
\affil{Physics Department and Tsinghua Center for Astrophysics (THCA), Tsinghua University, Beijing, 100084, People's Republic of China}

\author{Peng Wei}  
\affil{Xinjiang Astronomical Observatory, Chinese Academy of Sciences, Urumqi, Xinjiang 830011, People's Republic of China}

\author{Tianmeng Zhang} 
\affil{Key Laboratory of Optical Astronomy, National Astronomical Observatories, Chinese Academy of Sciences, Beijing 100012, People's Republic of China}

\author[0000-0002-5060-3673]{D. E. Reichart}  
\affiliation{Department of Physics and Astronomy, University of North Carolina at Chapel Hill, Chapel Hill, NC 27599, USA}

\author{Thomas G. Brink} 
\affil{Department of Astronomy, University of California, Berkeley, CA 94720-3411, USA}




\author{Curtis McCully}  
\affil{Las Cumbres Observatory, 6740 Cortona Drive Suite 102, Goleta, CA 93117-5575, USA}
\affil{Department of Physics, University of California, Santa Barbara, CA 93106-9530, USA}

\author[0000-0002-1125-9187]{Daichi Hiramatsu}  
\affil{Las Cumbres Observatory, 6740 Cortona Drive Suite 102, Goleta, CA 93117-5575, USA}
\affil{Department of Physics, University of California, Santa Barbara, CA 93106-9530, USA}

\author{Griffin Hosseinzadeh}  
\affil{Center for Astrophysics, Harvard \& Smithsonian, 60 Garden Street, Cambridge, MA 02138-1516, USA}
\affil{Las Cumbres Observatory, 6740 Cortona Drive Suite 102, Goleta, CA 9311-5575, USA}
\affil{Department of Physics, University of California, Santa Barbara, CA 93106-9530, USA}

\author{Benjamin T. Jeffers}  
\affil{Department of Astronomy, University of California, Berkeley, CA 94720-3411, USA}

\author{Timothy W. Ross} 
\affil{Department of Astronomy, University of California, Berkeley, CA 94720-3411, USA}

\author{Samantha Stegman}  
\affil{Department of Astronomy, University of California, Berkeley, CA 94720-3411, USA}
\affil{Department of Chemistry, University of Wisconsin, Madison, WI 53706, USA}

\author{Lifan Wang}  
\affil{George P. and Cynthia Woods Mitchell Institute for Fundamental Physics \& Astronomy, Texas A\&M University, Department of Physics and Astronomy, 4242 TAMU, College Station, TX 77843, USA}

\author{Jicheng Zhang} 
\affil{Physics Department and Tsinghua Center for Astrophysics (THCA), Tsinghua University, Beijing, 100084, People's Republic of China}

\author{Shuo Ma} 
\affil{Xinjiang Astronomical Observatory, Chinese Academy of Sciences, Urumqi, Xinjiang 830011, People's Republic of China}
\affil{School of Astronomy and Space Science, University of Chinese Academy of Sciences, Beijing 100049, People's Republic of China}

\begin{abstract}
We present extensive optical photometric and spectroscopic observations of the high-velocity (HV) Type Ia supernova (SN Ia) 2017fgc, covering the phase from $\sim$\,12\,d before to $\sim 389$\,d after maximum brightness. SN 2017fgc is similar to normal SNe~Ia, with an absolute peak magnitude of $M_{\rm max}^{B} \approx$ \mbmag\,mag and a post-peak decline of \mb\ = \dmvalue\ mag. Its peak bolometric luminosity is derived as \Lmax, corresponding to a $^{56}$Ni mass of \MniValue. The light curves of SN 2017fgc are found to exhibit excess emission in the $UBV$ bands in the early nebular phase and pronounced secondary shoulder/maximum features in the $RrIi$ bands. Its spectral evolution is similar to that of HV SNe~Ia, with a maximum-light \SiII\ velocity of \VSiII\ and a post-peak velocity gradient of $\sim$ \VSiIIDot. The \FeII\ and \MgII\ lines blended near 4300\,\Angst\ and the \FeII, \SiII, and \FeIII\ lines blended near 4800\,\Angst\ are obviously stronger than those of normal SNe~Ia. Inspecting a large sample reveals that the strength of the two blends in the spectra, and the secondary peak in the $i/r$-band light curves, are found to be positively correlated with the maximum-light \SiII\ velocity. Such correlations indicate that HV SNe~Ia may experience more complete burning in the ejecta and/or that their progenitors have higher metallicity. Examining the birthplace environment of SN 2017fgc suggests that it likely arose from a stellar environment with young and high-metallicity populations. 
\end{abstract}


\section{Introduction}

Type Ia supernovae (SNe~Ia) are widely believed to originate from thermonuclear runaway explosions of carbon-oxygen (CO) white dwarfs (WDs) in binary systems \citep[e.g.,][]{Nomoto1997,Hillebrandt200038,Maoz201452,Livio2018736,Soker2019}, and 
they have a typical absolute $V$-band peak magnitude of $\sim -19$\,mag \citep[e.g.,][]{Phillips1993,Perlmutter1999,wangxiaofeng2006ApJ}. One of the main applications of SNe~Ia is that they can be utilized as extragalactic distance indicators \citep[e.g.,][]{Riess1996,wangxiaofeng2005ApJL,Guy2005,Howell2006Nat,Howell2011NatCo,Burns2018,Scolnic2018859}, leading to the discovery of the accelerating expansion of the Universe \citep{Riess1998,Perlmutter1999}.

Two prevailing ideas for the progenitor systems are the double-degenerate 
(DD) scenario \citep{Webbink1984,Iben1984ApJS} and the single-degenerate (SD) scenario \citep{Whelan1973,Nomoto1984286,Podsiadlowski2008NewAR}. 
The former scenario involves the dynamical merger of two WDs with an accretion phase \citep{Rasio1994432,Yoon2007380,Pakmor2012747,Sato2015807} or the violent/third-body-induced collision of a binary WD \citep{Thompson2011741,Pakmor2012747,Raskin2014788,Sato2015807}, while explosion in the latter case is triggered by accretion onto a WD from its nondegenerate companion \citep{Whelan1973,Webbink1984,Iben1984ApJS}.
However, details of the progenitor systems and explosion mechanisms of SNe~Ia are still controversial \citep{wangxiaofeng2013Sci,Maoz201452,Jha2019NatAs,Han2020}. 
Some tentative evidence presented for the absence of companion stars in some SNe~Ia favors the DD scenario \citep{Gonz2012Natur,Schaefer2012Natur,Olling2015Natur,Tucker2019}, while the possible detections of circumstellar material (CSM) support the SD scenario for at least a portion of SNe~Ia \citep{Hamuy2003Natur,Wang2004,Aldering2006,Pastorello2007,Blondin2009,Sternberg2011Sci,Taddia2012,Dilday2012Sci,Silverman2013,Bochenek2018,wangxiaofeng2019ApJ}, though some theoretical studies show that the CSM could be also produced in the DD scenario \citep{Raskin2013772,Shen2013770,Levanon2017470}.

Observations show that $\sim70\%$ of SNe~Ia are members of the spectroscopically normal subclass \citep{Branch1993,LiWD2011}, while the others are 
classified as peculiar, including the overluminous SN 1991T-like \citep{Filippenko1992a,Ruiz1992,Filippenko1997}, the subluminous SN 1991bg-like \citep{Filippenko1992b,Leibundgut1993}, and the low-maximum-light velocity 
with low luminosity SN~Iax~2002cx-like \citep{Filippenko2003,LiWD2003,Foley2013767}. \citet{Benetti2005} divided normal SNe~Ia into three subclasses: high velocity gradient (HVG), low velocity gradient (LVG), and faint, 
according to the temporal velocity gradient of the \SiII\ line. Based on the equivalent width (EW) of \SiII\,$\lambda$6355 and \SiII\,$\lambda$5972 measured at $B$-band maximum, \citet{Branch2006118} proposed that SNe~Ia could be classified into core normal (CN), broad line (BL), cool (CL), and shallow silicon (SS) subgroups. According to the \SiII\,$\lambda$6355 
velocity measured at $B$-band maximum, \citet{wangxiaofeng2009ApJL} suggested that the ``Branch-normal'' SNe~Ia could be classified into normal-velocity (i.e., $v\,\textless\,12,000$\,\kms; NV) and high-velocity (i.e., $v \geq 12,000$\,\kms; HV) subclasses. Although various classification schemes have been proposed to classify SNe~Ia, there are overlaps between different classifications as demonstrated by different samples \citep{Blondin2012, Silverman2012c}. For example, HV SNe~Ia are usually found to have large velocity gradient and broad line profiles \citep{Barbon1989,Benetti2004,wangxiaofeng2009ApJL,Yamanaka200961,Silverman2012b}. Moreover, after the analysis of the birthplace environments of SNe~Ia in their host galaxies,  \citet{wangxiaofeng2013Sci} suggest that the HV and NV subclasses of SNe~Ia may come from progenitor systems with different metallicities.

A recent study by \citet{Pan2015446} and \citet{LiWX2021906} suggest that HV~SNe~Ia tend to reside in massive galaxies and likely have metal-rich progenitor environments. 
Through an investigation of a large set of the host galaxies, \citet{Pan2020895} found that HV~SNe might arise in massive host galaxies with metal-rich environments. Based on an analysis of \NaI\ absorption lines in the spectra and the late-time light curves in the $B$ and $V$ bands, \citet{wangxiaofeng2019ApJ} show that HV~SNe~Ia likely have more abundant circumstellar dust around their progenitors, and they may originate from progenitor systems with nondegenerate companions. 

SN 2017fgc is a fast-expanding SN~Ia that exploded in the nearby shell galaxy NGC 474 at a distance of $29.51 \pm 2.09$ Mpc \citep{Cappellari2011413}. Note that the host is a lenticular galaxy with prominent gas shell and bridge structures at its outskirts, suggestive of the merging process \citep{Lim2017835,Alabi2020497,Fensch2020644}. \citet{Burgaz2021502} studied the photometric properties of this SN and proposed the presence of a prominent $RI$-band secondary shoulder/maximum. Here we present extensive 
photometric and spectroscopic observations of SN 2017fgc, and we report an additional discrepancy between the NV and HV subclasses in a well-observed sample of SNe~Ia.

In this paper, the optical observations and data reduction are presented in Section 2. Section 3 discusses the light curve, color curves, and quasibolometric light curve, while Section 4 presents the spectroscopic evolution. The properties of SN 2017fgc and its host galaxy are discussed in Section 5. We summarize in Section 6.

\section{Observations and Data Reduction}

\subsection{Discovery and Host Galaxy}

SN 2017fgc was discovered at $\alpha =01^h20^m14^s.440$, $\delta = 03^{\circ}24\arcmin09\arcsec.96$ (J2000) on 2017 July 9.29 (UT dates are adopted throughout this paper) by the Distance Less Than 40\,Mpc (DLT40; \citealp{DLT40ref}) survey at $r = 17.32$\,mag \citep{2017fgcDiscover}. At that time, the DLT40 survey operated with a $\sim24$\,hr cadence, and the last nondetection of the SN was on 2018 July 8.29 with a limiting magnitude of $r\approx19.5$\,mag. A spectrum taken $\sim 1.4$\,hr after the discovery classified it as a normal SN~Ia \citep{2017fgcClass}. The host galaxy of SN 2017fgc is NGC 474 with a redshift of $z = 0.00772 \pm 0.00002$ \citep{Hern2011}, which corresponds to a distance modulus of $\mu = 
32.51 \pm 0.11$\,mag assuming a Hubble constant of 73.5\,\kms\,Mpc$^{-1}$ 
\citep{Riess2018}.

\subsection{Photometry}

\begin{figure*}[ht]
\centering
\includegraphics[angle=0,width=172mm]{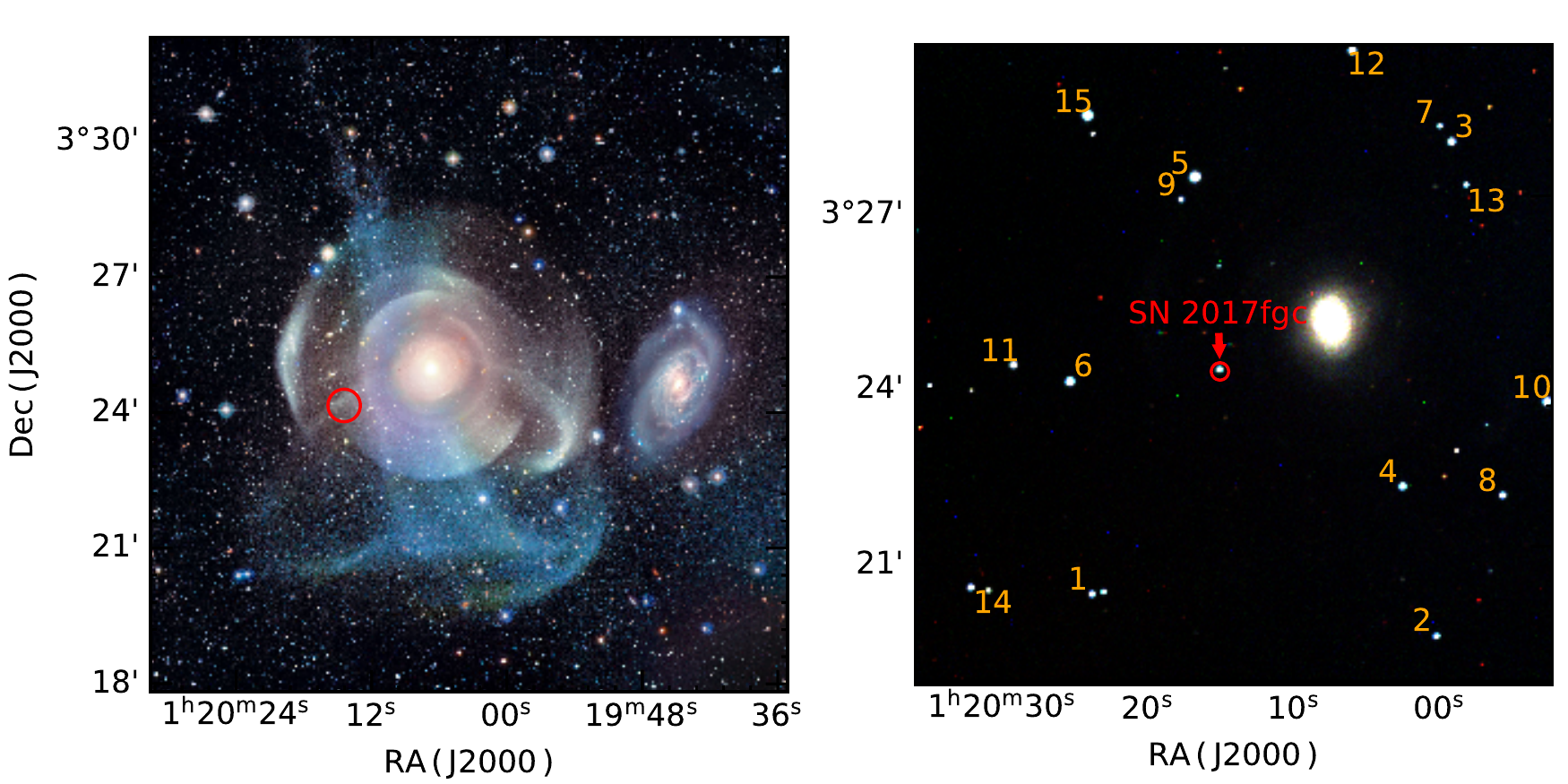}
\caption{The left panel shows a color image which is synthesized from observations in $gr$ bands obtained by the CFHT before the discovery of SN 2017fgc. The red circle marks the position of SN 2017fgc and the gas bridge can be clearly seen here. The right panel shows a color image synthesized from TNT observations in the $gri$ bands after the explosion of SN 2017fgc, and the SN is marked with a red circle while the reference stars are labeled by numbers. \label{fobsimg}}
\end{figure*}

Our photometric observations of SN 2017fgc were obtained with several telescopes, including the 0.8\,m Tsinghua-NAOC telescope (TNT; \citealp{Huang2012}), the Las Cumbres Observatory (LCO) Telescope network \citep{Brown2010}, the 0.76\,m Katzman Automatic Imaging Telescope (KAIT) at Lick Observatory \citep{Filippenko1999,Filippenko2001}, and the 1\,m Nickel reflector\footnote{https://www.ucolick.org/public/telescopes/nickel.html} at Lick Observatory. The TNT and Nickel telescope monitored SN~2017fgc in the 
$BVRI$ bands, KAIT observed it in the $BVRI$ and Clear bands, and the LCO 
1\,m telescopes sampled the light curves in the $UBVgri$ bands. Figure \ref{fobsimg} shows color images of SN 2017fgc synthesized from observations in $gri$ bands. 
The left panel shows a color image is synthesized from CFHT\footnote{https://www.cfht.hawaii.edu/HawaiianStarlight/images.html} observations in $gr$ bands processed by J.-C. Cuillandre and G. Anselmi, based on data from the MATLAS program \citep{Duc2015446}, while the right panel shows a color image systhesized from TNT observations.

For photometric images obtained from the LCO during the Global Supernova Project, $lcogtsnpipe$ \citep{lcogtpipe} was employed for image reduction. The point-spread functon (PSF) implemented in $Photutils$ \citep{Bradley2020} was utilized to extract instrumental magnitudes of SN 2017fgc from images obtained by LCO. Photometric images from KAIT and Nickel were reduced with LOSSPhotPypeline\footnote{\url https://github.com/benstahl92/LOSSPhotPypeline} \citep{Stahl2019LOSS,Stahl2020depSIP}. The DLT40 images 
were reduced using a dedicated difference-imaging pipeline \citep{Tartaglia2018853} which calibrates the Open filter data to the $r$ band utilizing the AAVSO Photometric All-Sky Survey (APASS\footnote{\url https://www.aavso.org/apass}) catalog. PSF photometry was performed on the difference images. As SN 2017fgc was located far away ($\sim 18.90 \pm 0.01$\,kpc) from the center of its host galaxy, we did not apply image-subtraction techniques for the photometry.

\begin{figure}[ht]
\centering
\includegraphics[angle=0,width=86mm]{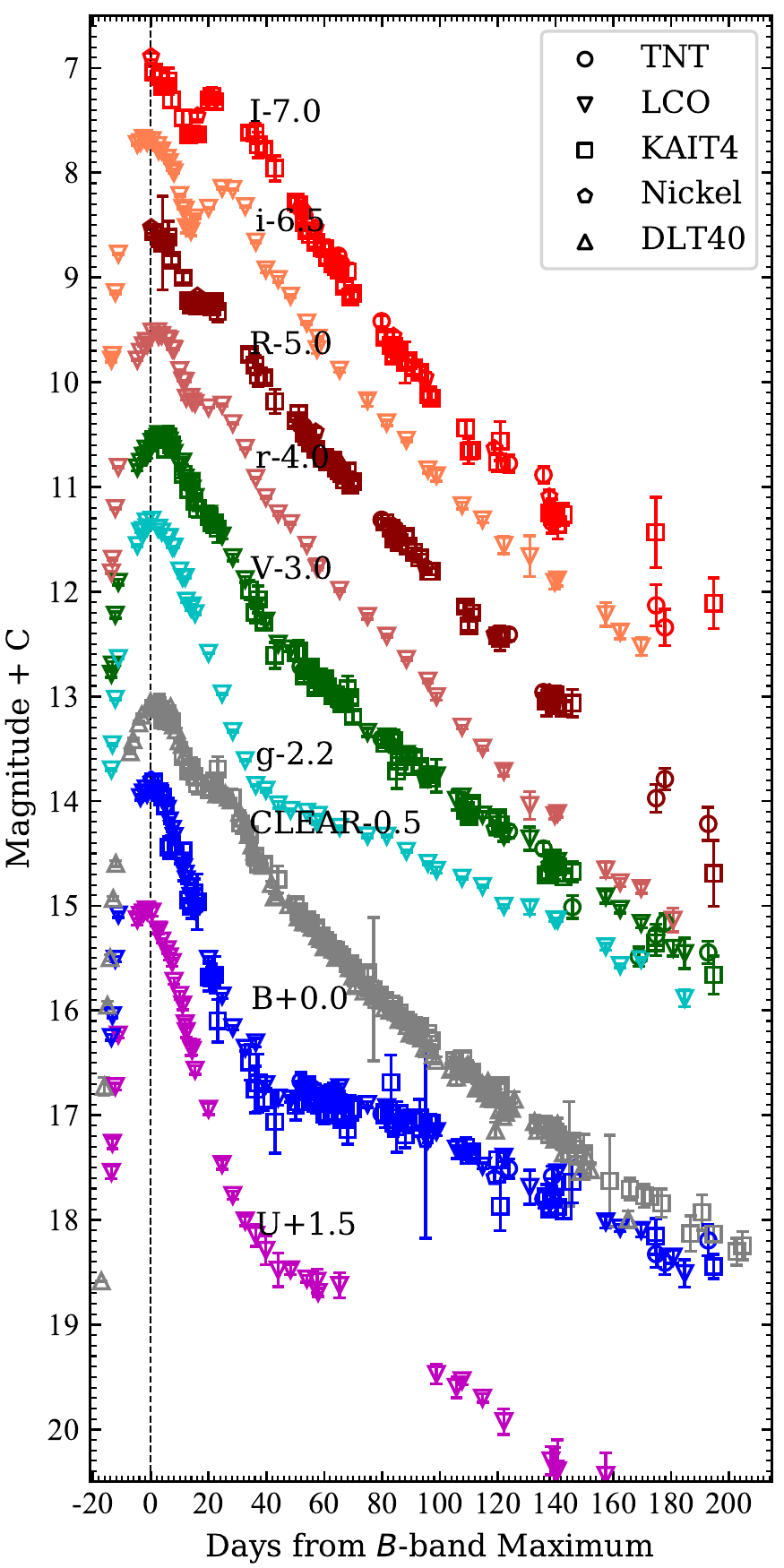}
\caption{The optical light curves of SN 2017fgc. The different colors represent different observation bands, while data from different telescopes/instruments are indicated with different symbols. The vertical dashed line represents the time of the $B$-band maximum, relative to which all times are given in this paper. The light curves have been shifted vertically for clarity. \label{fgclcv}}
\end{figure}

The color-transformation method introduced by \citet{Jordi2006} was employed to convert $gri$ magnitudes from SDSS Dr12 \citep{Alam2015} to Landolt $UBVRI$ \citep{Landolt1992} magnitudes. The local standard stars with SDSS $gri$ magnitudes and the transformed $UBVRI$ magnitudes are listed in 
Table \ref{tab:standards}. The instrumental magnitudes obtained from TNT and KAIT in $BVRI$ are calibrated to the Johnson $UBVRI$ system \citep{Landolt1992,Stetson2000}. The unfiltered magnitudes from KAIT are calibrated to the standard Landolt $R$-band magnitudes, with a typical uncertainty 
of about 0.2--0.3\,mag \citep{LiWD2003, Zheng2017b}. The LCO instrumental 
magnitudes in $UBV$ are calibrated to the Johnson system \citep{Landolt1992,Stetson2000}, while the $gri$ magnitudes are calibrated using the SDSS 
Dr12 catalog \citep{Alam2015}. The final light curves are shown in Figure 
\ref{fgclcv}, and the flux-calibrated magnitudes are listed in Table \ref{fgc:gndphot}.

\subsection{Spectroscopy}

A total of 38 low-resolution spectra were obtained for SN 2017fgc with different instruments, including the FLOYDS spectrographs mounted on the LCO-2\,m Faulkes Telescope North and South (FTN and FTS; \citealp{Brown2013LCO2m, Floyds}), the BFOSC mounted on the Xinglong 2.16\,m telescope (XLT; \citealp{Jiang1999XLT, ZhangJC2016XLT, Fan2016XLT}), the YFOSC on the Lijiang 2.4\,m telescope (LJT; \citealp{Chen2001LJT0, WangCJ2019LJT1}) of Yunnan Astronomical Observatories, and the Kast spectrograph on the Lick 3\,m Shane telescope \citep{Fili1986Lick3m, Vogt1987Lick3m, Stahl2020Spec}. Three additional spectra of SN 2017fgc were obtained with XSHOOTER \citep{Vernet2011536} mounted on the ESO Very Large Telescope (VLT), at $t\approx+149.7$\,d, +383.9\,d, and 388.9\,d during ESO programs 0101.D-0242(A) and 0101.D-0443(A) \citep{Graur20204}. The journal of spectroscopic observations is presented in Table \ref{tab:spectra}.

Standard IRAF\footnote{IRAF is distributed by NOAO, which is operated by AURA, Inc., under cooperative agreement with the U.S. National Science Foundation (NSF).} routines were used to reduce the spectra. Spectrophotometric standard stars observed at an airmass comparable to the target on the same night were used for flux calibration. The extinction curves at various observatories were utilized to correct for atmospheric extinction, and spectra of the standard stars were used to eliminate the telluric absorption lines.

\section{Light Curves}

\subsection{Optical Light Curves}

Figure \ref{fgclcv} shows the optical light curves of SN 2017fgc, covering the phases from about two weeks before to over 200\,d after $B$-band maximum light. Overall, they are similar to those of normal SNe~Ia, characterized by a prominent shoulder or secondary maximum in the $R/r$ and $I/i$ bands. A small shoulder may also be detectable in the $V$ and Clear bands. Applying a polynomial fit to the $B$-band light curves around maximum 
light yields a peak value of \mB\,mag on MJD = 57959.4 (2017 July 25.4 UT). The $V$-band light curve reached its peak of \mV\,mag on MJD = 57962.5, $\sim 3.1$\,d after the $B$-band peak. The postpeak $B$-band decline rate \mb\ is measured as \dmvalue\,mag, and the color stretch \citep{Burns2011,Burns2014} is determined to be \sBV = \sbvalue.

\begin{figure*}[ht]
\centering
\includegraphics[angle=0,width=172mm]{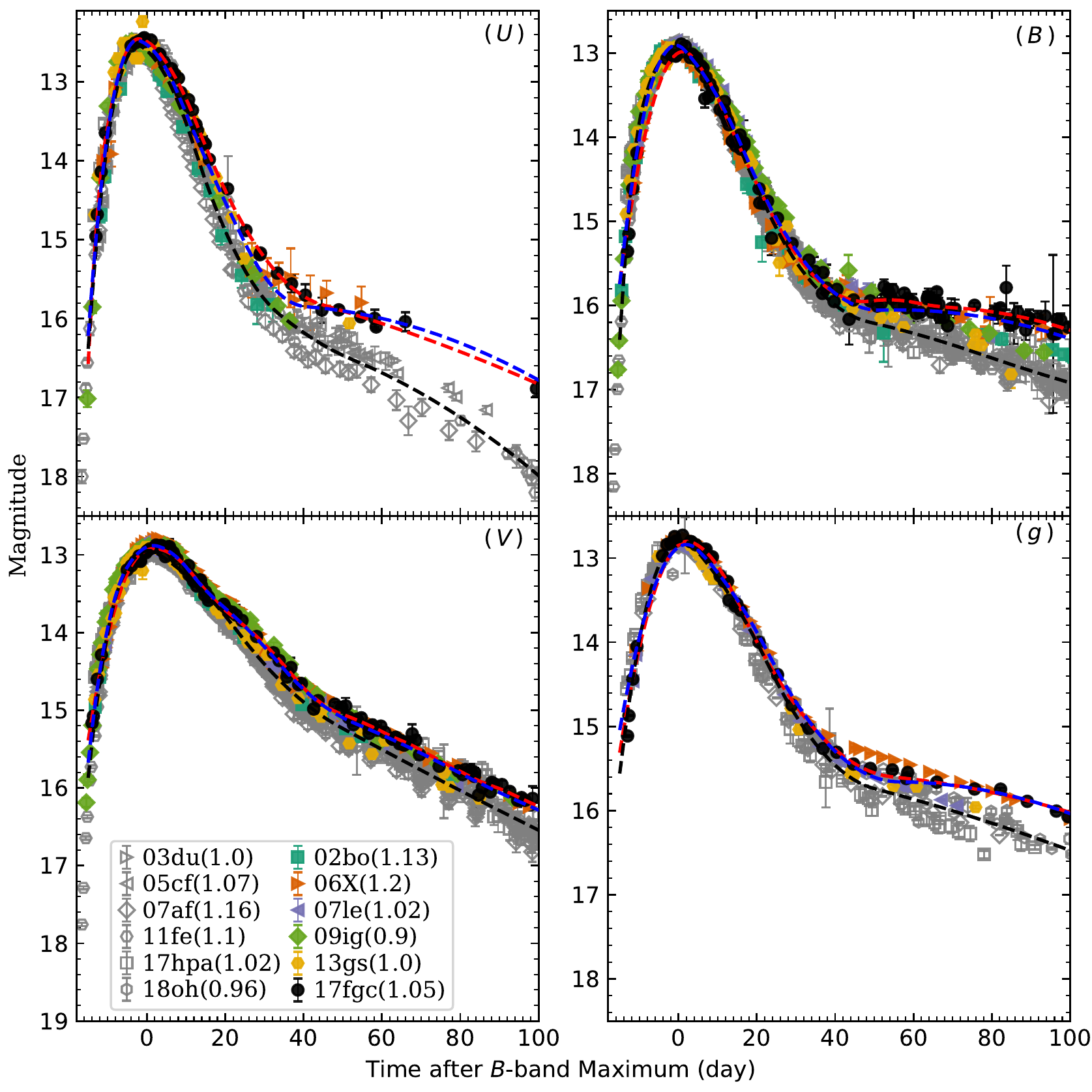}
\caption{Comparison of the optical light curves (in the $UBVg$ bands) of SN 2017fgc with those of other well-observed SNe~Ia having similar decline rates. The rest-frame light curves of the comparison SNe~Ia have been normalized to match the peak magnitudes of SN 2017fgc. No k-corrections have been applied to the light curves, as both SN 2017fgc and the comparison sample have a redshift less than 0.01. The NV~SNe~Ia are marked with open symbols while the HV~SNe are shown with filled symbols. The black dashed lines represents the polynomial fit to the NV~SNe~Ia, while the blue dashed lines represents the case for HV SNe Ia. The red dashed lines represents the best-fit for the light curves of SN 2017fgc. \label{fgcBlueLC}}
\end{figure*}

\begin{figure*}[ht]
\centering
\includegraphics[angle=0,width=172mm]{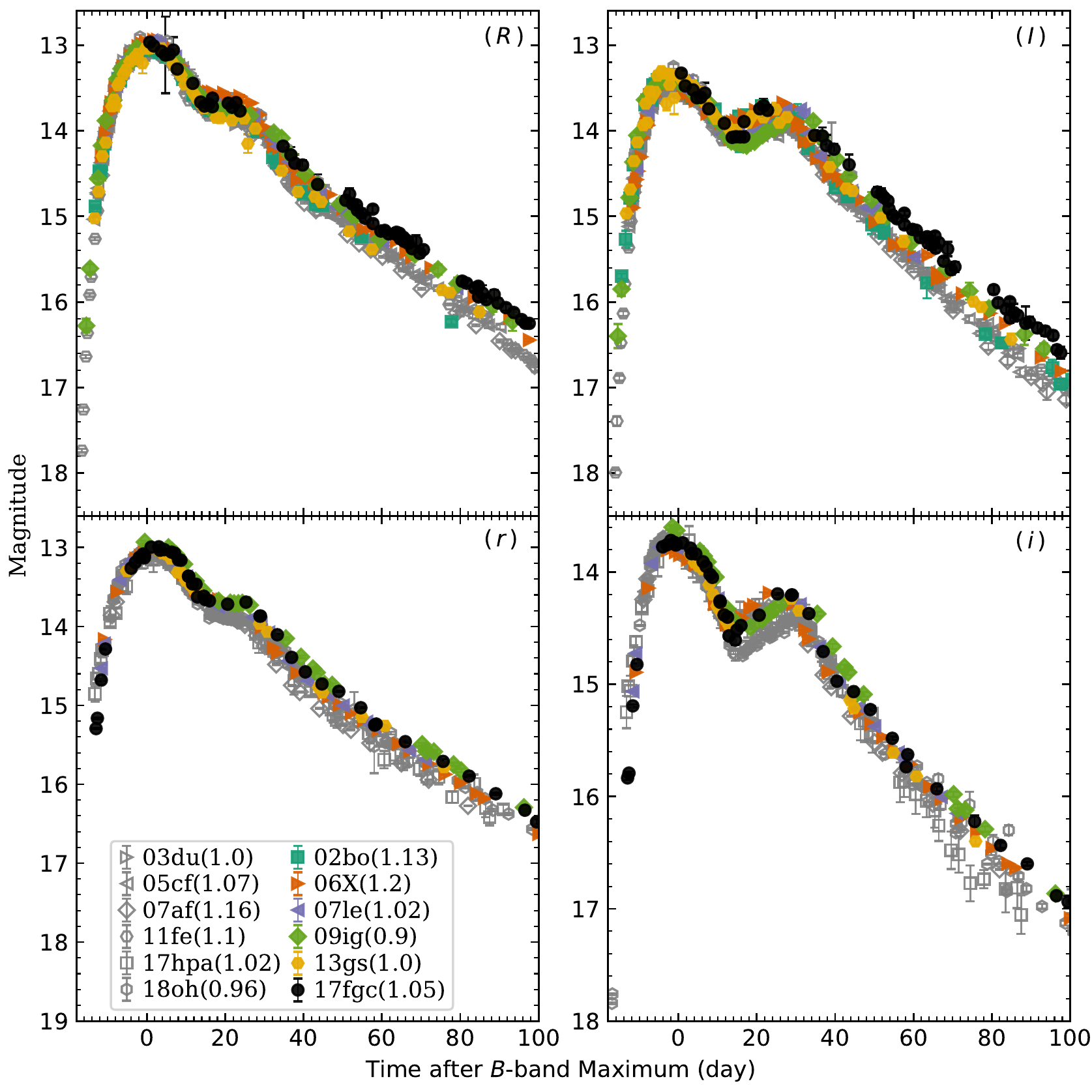}
\caption{Comparison of the optical light curves (in the $RrIi$ bands) of SN 2017fgc with those of other well-observed SNe~Ia having similar decline rates. The light curves of the comparison SNe~Ia have been normalized to match the peak magnitudes of SN 2017fgc. The symbols are the same as in 
Figure \ref{fgcBlueLC}. \label{fgcRedLC}}
\end{figure*}

In Figures \ref{fgcBlueLC} and \ref{fgcRedLC} the $UBVg$ and $RrIi$ light 
curves of SN 2017fgc are compared with those of well-observed normal SNe~Ia having similar \mb, including SNe~2002bo \citep{Krisciunas2004, Benetti2004}, 2003du \citep{Stanishev2007469}, 2005cf \citep{wangxiaofeng2009ApJ}, 2006X \citep{wangxiaofeng2008ApJ}, 2007af \citep{Stritzinger2011142}, 
2007le \citep{Ganeshalingam2010190, Hicken2012200}, 2009ig \citep{Marion2013777}, 2011fe \citep{Munari2013, ZhangKC2016}, 2013gs \citep{zhangtianmeng2019ApJ}, 2017hpa \citep{Zeng2021909}, and 2018oh \citep{liwenxiong2019a}. One can see that the light curves of SN 2017fgc are generally similar to those of comparison SNe~Ia near the $B$-band maximum. We notice, however, that SN 2017fgc seems to have brighter tails in both $U$ and $B$ relative to those of NV~SNe~Ia, consistent with the tendency that HV~SNe~Ia have flatter evolution at $t\approx1$-3 months after maximum light (e.g., \citealp{wangxiaofeng2008ApJ,wangxiaofeng2019ApJ}). For example, the magnitude decline measured within 60\,d after peak brightness is $3.57 \pm 0.08$\,mag in $U$ and $2.96 \pm 0.08$\,mag in $B$, similar to the evolution of the well-known HV~SN~2006X. Figure \ref{fgcRedLC} 
shows a comparison of the $RrIi$ light curves. One can see that SN 2017fgc has a more significant secondary shoulder or peak in the $R/r$ and $I/i$ bands than the normal counterparts. 

\subsection{Reddening}

The line-of-sight Galactic extinction of SN 2017fgc is estimated to be $A_{V} = 0.094$\,mag \citep{sfd1998,sfd2011}. Adopting $R_{V} = 3.1$ \citep{ccm1989}, this corresponds to a color excess of $E(B-V)_{\rm Gal} = 
0.030$\,mag. After removal of the Galatic extinction, the $B-V$ color of SN 2017fgc is estimated to be $-0.03 \pm 0.02$\,mag at $t = 0$\,d and $1.62 \pm 0.10$\,mag at $t = +35$\,d relative to the $B$-band maximum, consistent with typical values of normal SNe~Ia \citep{Phillips1999,wangxiaofeng2009ApJL}.

\begin{figure*}[ht]
\centering
\includegraphics[angle=0,width=172mm]{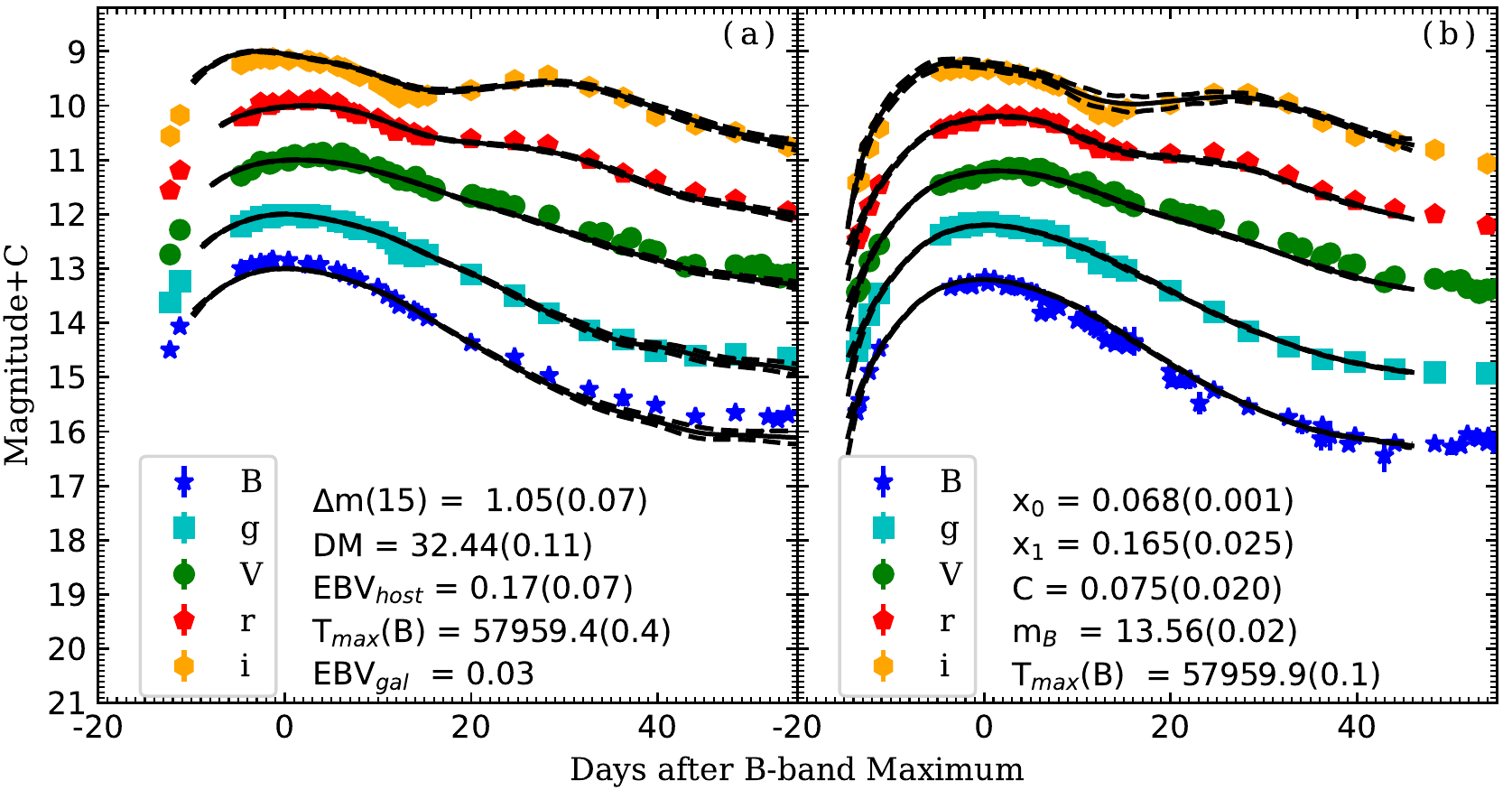}
\caption{Best-fit light-curve model from SNooPy2 (left panel) and SALT 2.4 (right panel). The light curves are shifted vertically for clarity. The 
dashed lines represent the 1$\sigma$ uncertainty of the best-fit light-curve templates. \label{fsnpymodel}}
\end{figure*}

SuperNovae in object-oriented Python (SNooPy2; \citealp{Burns2011, Burns2014}) has also been employed to fit the multiband light curves of SN 2017fgc, and the model fitting results are shown in Figure \ref{fsnpymodel}(a). We adopt the $EBV$ model with $st$-type and estimate the reddening due 
to the host galaxy as $E(B-V)_{\rm host}$ = \ebvalue\,mag, suggesting an insignificant host-galaxy reddening for SN 2017fgc. This is consistent with the absence of \NaI~D absorption lines in the optical spectra of SN 2017fgc. 

\subsection{Color Curves}

\begin{figure*}[ht]
\centering
\includegraphics[angle=0,width=172mm]{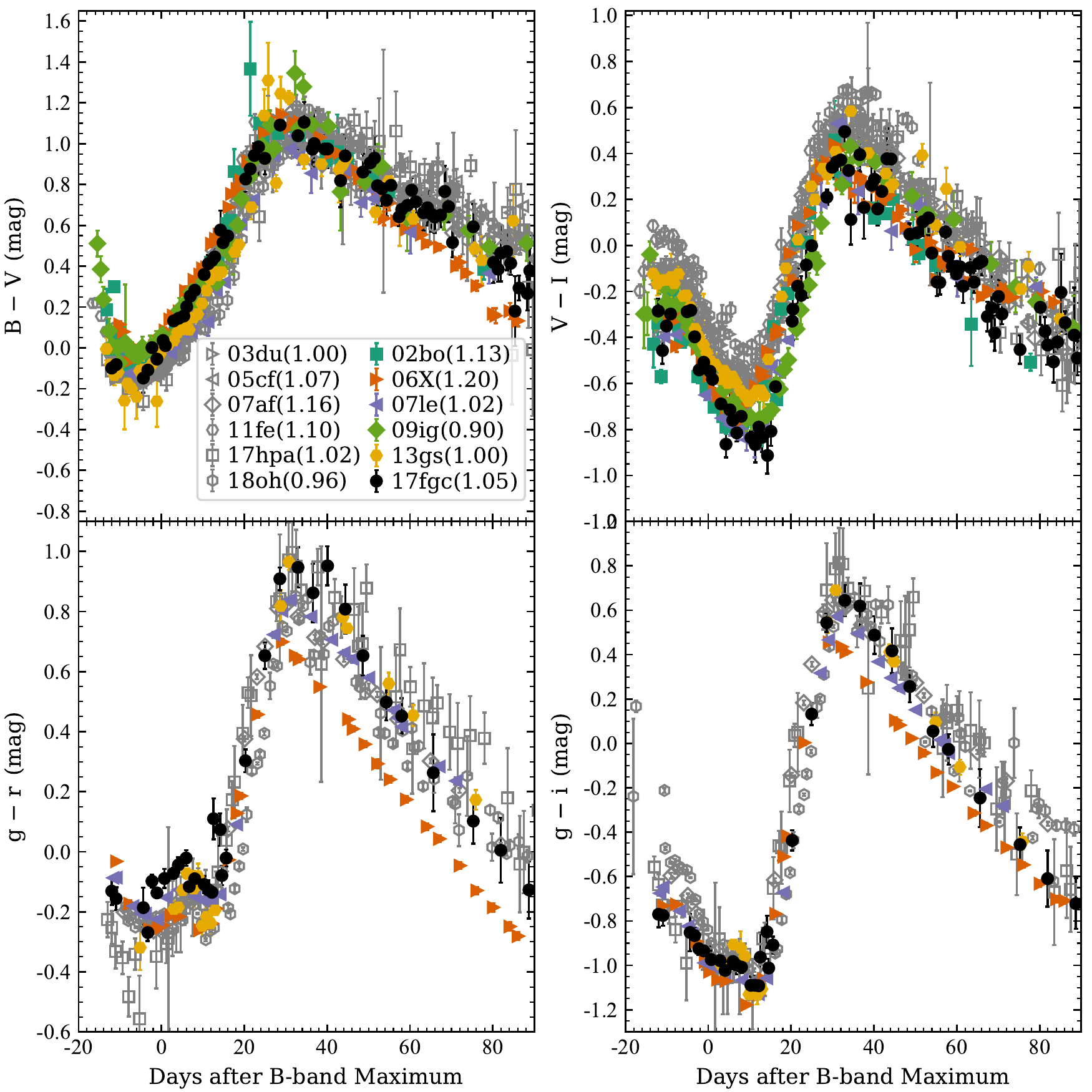}
\caption{The $B-V$, $V-I$, $g-r$, and $g-i$ color curves of SN 2017fgc are compared with those of SNe~2002bo, 2003du, 2005cf, 2006X, 2007af, 2009ig, 2011fe, 2013gs, 2017hpa, and 2018oh. All color curves, including those of SN 2017fgc, have been dereddened using SNooPy2 \citep{Burns2011,Burns2014}. \label{fcolorcomp}}
\end{figure*}

Figure \ref{fcolorcomp} shows the color evolution of SN 2017fgc compared with that of several well-observed SNe~Ia having similar \mb. 
At early times, both the $B-V$ and $g-r$ color curves evolve toward the blue until reaching the blue peak at $t \approx 0$\,d; then they evolve redward and reach the red peak at $t \approx 35$\,d. 
The $B-V$ and $V-I$ color curves of SN 2017fgc show close resemblances to those of SN 2002bo, SN 2006X, and SN 2009ig, which all belong to the subclass of HV~SN~Ia \citep{Krisciunas2004,Benetti2004,wangxiaofeng2008ApJ,Silverman2012b, Hicken2012200}. However, none of the comparison SNe Ia look perfectly the same as SN 2017fgc.
While evolving redwards, the $g-r$ color curves of SN 2017fgc and other comparison SNe Ia show a clear jump at $t \sim
 $\,1 week after maximum light, which is similar to that of the $V-R$ color curves at similar phase \citep{Cartier2014789,Guti2016590,zhangtianmeng2019ApJ}.
Both the $V-I$ and $g-i$ color curves show an evolution from red to blue until $t \approx 15$\,d; then they evolve redward and reach the red peak at $t \approx 35$\,d. 
Like other HV SNe Ia, SN 2017fgc also exhibited an overall bluer $V-I$ color in comparison with the NV counterparts.
The prominent \CaII\ near-infrared (NIR) absorption features shown in the spectra of HV~SNe~Ia are likely responsible for their bluer $V-I$ colors.

\subsection{First-Light Time}

\begin{figure}[ht]
\centering
\includegraphics[angle=0,width=86mm]{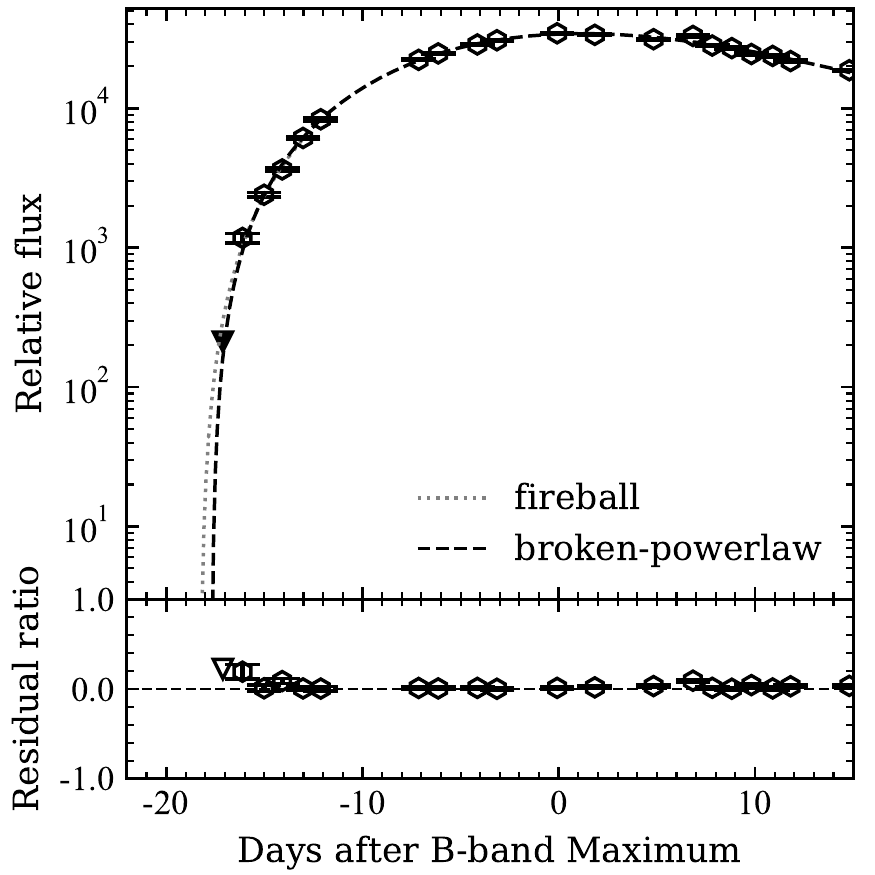}
\caption{Fit to the observed clear band light curve using the ideal fireball model \citep{Riess1999118} and the analytic function introduced by \citet{Zheng2017c}. The black triangle represents the earliest limit-magnitude from the DLT40 survey which is $\sim$\,0.6 days after the fitted first-light time.\label{ftrise}}
\end{figure}

The Distance Less Than 40 (DLT40) Mpc survey intensely observed SN2017fgc at extremely early times, using a clear band (calibrated to $R$-band). The photometric data are listed in Table \ref{dlt40phot}. Both the expanding fireball model from \citet{Riess1999118} and the broken power-law model from \citet{Zheng2018858} are adopted to fit the early light curve of SN 2017fgc (as shown in Figure \ref{ftrise}). 
The observed data at $t$\,$\leq$\,-10 days from the $R$-band peak are adopted to fit the fireball model while those obtained at $t$\,$\leq$\,+15 days are used for the broken power-law model fitting, and the estimated first-light time (FLT) of the light curve are 57941.1$\pm$0.3\,d and 57941.7$\pm$0.7\,d, respectively. 
The results of these two estimations are in a good agreement. The average fitted first-light time is estimated as 57941.4$\pm$0.4\,d, hence, the rise time of SN 2017fgc is estimated as 18.0$\pm$0.4\,d. The multicolor observations of SN 2017fgc is conducted at $\sim$6 days after the FLT, $\sim$12 days before the $B$-band maximum. 
A flux excess seems to exsit for the first detection point, i.e., brighter than the fitting result by $\sim$0.23$\pm$0.07 mag (see the bottom panel of Fig. \ref{ftrise}). 
This excess flux detected in the early phase could be due to interaction of SN ejecta with nondegenerate companion \citep{Kasen2010}, surrounding CSM \citep{Gerardy2004607,Piro2016826}, or even the radioactive decay of nickel synthesized on the surface of the exploding WD \citep{Piro2016826,Noebauer2017472}.
A further early-time light curves analysis including SN 2017fgc will be presented as part of a larger study (Burke et al. in preparation).

\begin{figure}[ht]
\centering
\includegraphics[angle=0,width=86mm]{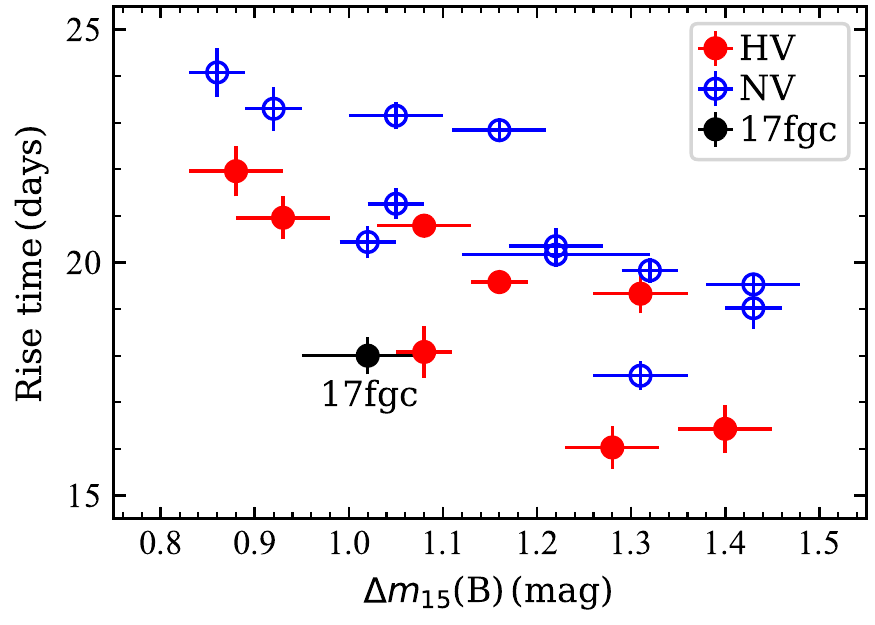}
\caption{$R$-band rise times of several well-observed SNe Ia are plotted against the light curve decline rate \mb. The data are taken from \citet{ZhangTM2010}, and SN 2017fgc is overplotted in black dot. The HV SNe Ia are plotted as red dots while the NV ones are blue open circles.\label{frtrise}}
\end{figure}

According to \citet{ZhangTM2010}, the rise time of the $r$-band light curve of SNe Ia show an anticorrelation with \mb\ (see Figure \ref{frtrise}). For given \mb, the rise time of NV SNe Ia seems to be on average longer than that of the HV ones. A similar trend has been reported in other studies (e.g., \citealp{Pignata2008388,Ganeshalingam2011416,Zheng2017c}). As shown in Figure \ref{frtrise}, SN 2017fgc also is in line with this trend. 
As the $R$-band and clear-band light curves have similar light curve shapes and magnitudes, we assume that they have similar rise time. The $R$-band rise time of SN 2017fgc is thus estimated as 18.0$\pm$0.4 day by fitting its clear-band light curve. Given that the $R$-band light curve of SN 2017fgc reached its peak ($\sim$57961.3\,$\pm$\,0.2) at $\sim\,$1.9\,$\pm$\,0.4 days later than that of the $B$-band ($\sim$57959.4\,$\pm$\,0.4), the $B$-band rise time can be thus inferred as 16.1\,$\pm$\,0.4\,d assuming that the SN photons in these two bands diffuse out simultaneously. 
The $B$-band rise time speculated for SN 2017fgc is also obviously shorter than the mean value of SNe Ia \citep{Zheng2017c}.
Compared to those of NV SNe Ia, the relatively shorter rise time seen in HV ones might be related to that their ejecta become optically thin at a faster pace because of more rapid expansion \citep{ZhangTM2010}.


\subsection{Quasi-Bolometric Light Curve}

According to \citet{Tully2013146}, based on the Tully-Fisher relation, the distance modulus measured for the host galaxy NGC 474 is $32.35 \pm 0.14$\,mag. From fitting to the multicolor light curves of SN 2017fgc, SNooPy2 gives an average $\mu = 32.44 \pm 0.11$\,mag (see left panel of Fig.~\ref{fsnpymodel}), while SALT 2.4 \citep{Guy2010523,Betoule2014568} gives $\mu = 32.39 \pm 0.07$\,mag (see right panel of Fig.~\ref{fsnpymodel}). These three estimates agree well with each other. The average value of 
\distmd\,mag is thus adopted in the following analysis. Assuming $R_{V} = 
3.1$, the absolute $B$-band peak magnitude of SN 2017fgc is estimated to be $M_{\rm max}(B)$ = \mbmag\,mag after correcting for both Galactic and host-galaxy extinction, which agrees well with that of normal SNe~Ia (i.e., $M_{\rm max}(B) \approx -19.3$\,mag  for an SN~Ia with \mb\ $\approx 1.1$\,mag; \citealp{Phillips1999,wangxiaofeng2009ApJL}).

Following the procedure used for SN 2018oh \citep{liwenxiong2019a}, SNooPy2 is employed to establish the spectral energy distribution (SED) and thus the quasibolometric light curve of SN 2017fgc based on the $U$, $B$, $g$, $V$, $R$, $r$, $I$, and $i$ light curves. 
According to \citet{wangxiaofeng2009ApJ} and \citet{ZhangKC2016}, the UV/optical ratio of SN 2005cf and SN 2011fe (Normal SNe Ia) are measured as 0.095 and 0.085, while the NIR/optical ratio of them are measured as 0.058 and 0.059, respectively. 
The converted ratios relative to the bolometric luminosity are 0.082 and 0.074 respectively for SN 2005cf and SN 2011fe in UV bands, while those ratios in NIR bands are 0.050 and 0.052, respectively. Based on the SED of SN 2009ig \citep{Marion2013777,Chakradhari2019487} near the maximum light, the UV/NIR to bolometric ratio of this HV SN Ia are estimated as 0.058$\pm$0.010 and 0.054$\pm$0.019, respectively. Assuming that the average ultraviolet (UV) and NIR contributions are 7\% and 5\% for SN 2017fgc, the maximum-light luminosity is estimated as $L_{\rm peak}$ = \Lmax, reached at $\sim 0.96$\,d prior to the $B$-band maximum.
This peak luminosity is larger than that of SN 2011fe ($\sim 1.13 \times 10^{43}$\,\ergs; \citealp{ZhangKC2016}) but smaller than that of SN 2018oh ($\sim 1.49 \times 10^{43}$\,\ergs; \citealp{liwenxiong2019a}).

\begin{figure}[ht]
\centering
\includegraphics[angle=0,width=86mm]{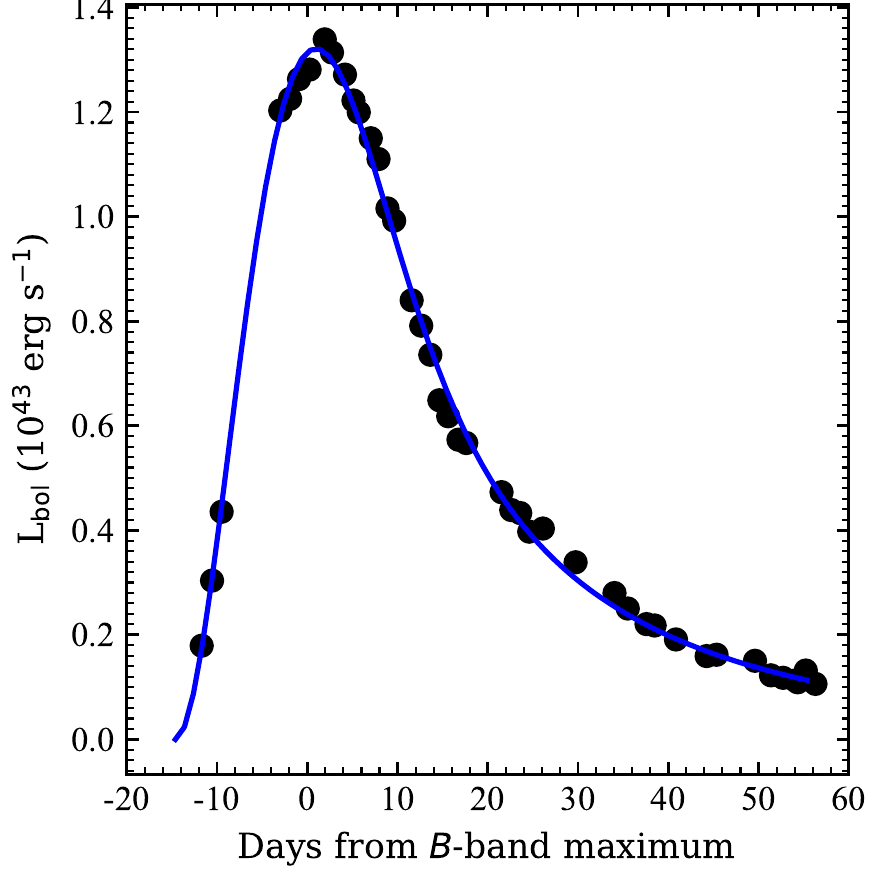}
\caption{Quasi-bolometric light curve (dots) with an Arnett (1982) radiation diffusion model (blue curve).\label{fgcBolometric}}
\end{figure}

The Minim Code \citep{Chatz2013}, utilizing a modified radiation diffusion model of Arnett \citep{Arnett1982, Chatz2012,liwenxiong2019a}, is used to fit the quasibolometric light curve with a constant-opacity approximation (see Figure \ref{fgcBolometric}). This fitting allows us to derive the following parameters, including 
the ``first light'' $t_{0}$, the model timescale of the light curve $t_{lc}$, the leaking timescale of gamma rays $t_{\gamma}$, and the radioactive \Nifs\ ejecta mass \Mni\ (see \citealp{Chatz2012} for details), with the respective values being \tzero, \tlcv\,d, \tgama\,d, and \MniValue. 
The dark phase was presented as the delay between the explosion and the emergence of the radio-activity-powered light curve in several models of SNe Ia \citep{Piro2013}. Assuming the initial diffusion wave reached the surface at $t$\,=\,0, the framework of Arnett model does not take a dark phase into account \citep{liwenxiong2019a}. \citet{Piro2016826} proposed that the length of the dark phase could be less than 2 days. The first light $t_{0}$ from the Arnett model is $\sim 2.1$\,d later than that estimated from the fireball model or the broken-power-law model. Considering the model-dependent uncertainties in the estimation, we interpreted that the difference in the two estimated starting moments could be related to the dark phase in SN 2017fgc \citep{Piro2016826,liwenxiong2019a}. The first light time of  57941.4$\pm$0.4\,d is used in the following analysis.


\section{Optical Spectra}

\subsection{Temporal Evolution of the Spectra}

\begin{figure}[ht]
\centering
\includegraphics[angle=0,width=86mm]{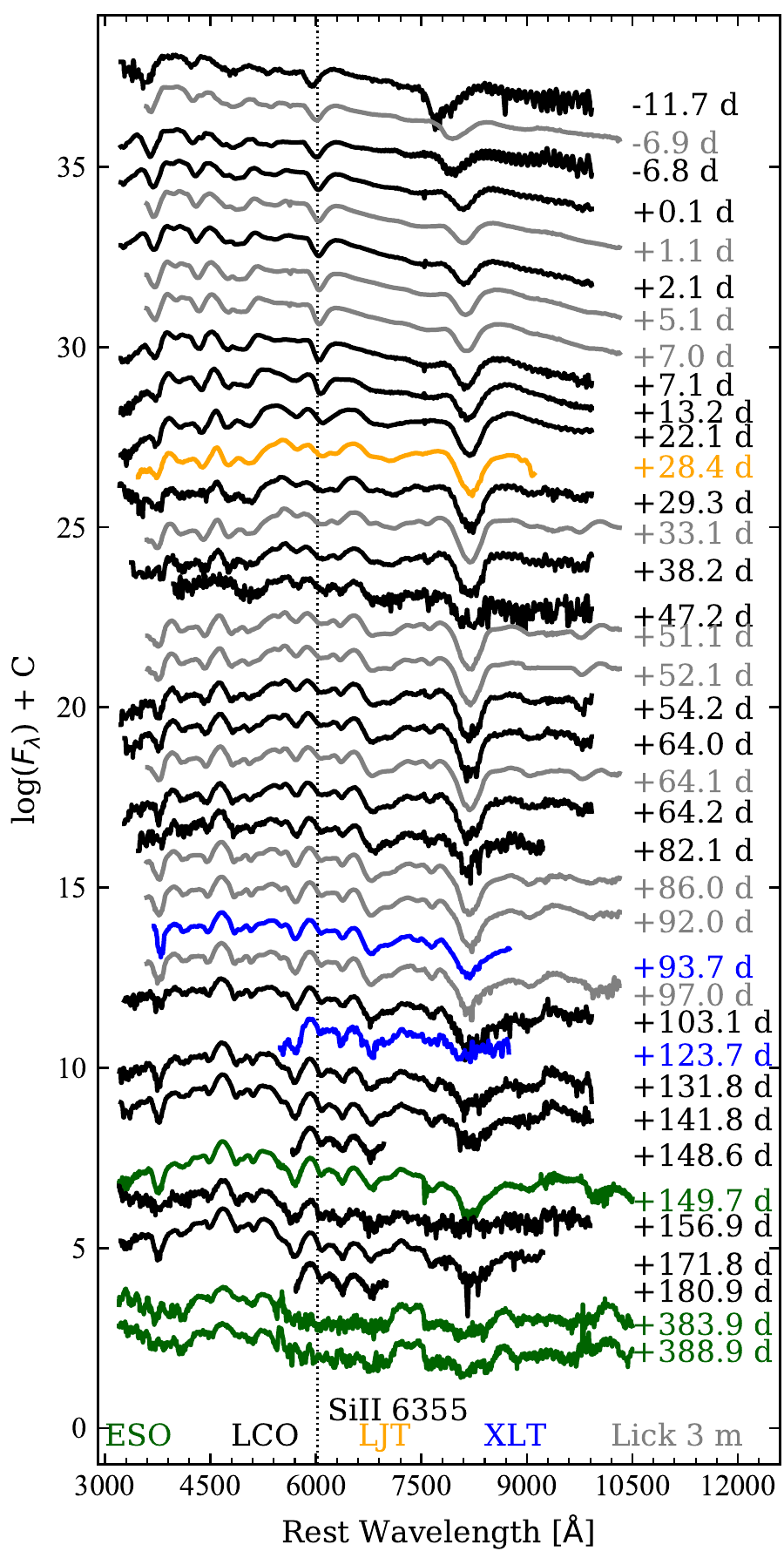}
\caption{Optical spectral evolution of SN 2017fgc. All of the spectra have been corrected for the redshift of the host galaxy and reddening. The epochs shown on the right side represent the phases in days relative to $B$-band maximum light. The dashed line marks the center of the \SiII~$\lambda$6355 line profile at +0.04\,d from $B$-band maximum. The colors of the spectra represent data from different instruments. \label{fspecevolution}}
\end{figure}

The optical spectral evolution of SN 2017fgc is displayed in Figure \ref{fspecevolution}. The early-time spectra are characterized by prominent absorption lines of intermediate-mass elements (IMEs) and ionized iron-group elements (IGEs), including \CaII\, H\&K, \FeII\,$\lambda\lambda$4404, 5018, \MgII\,$\lambda$4481, \SiII\,$\lambda$5051, \FeIII\,$\lambda$5129, \SII\,$\lambda\lambda$5468, 5654, \SiII\,$\lambda$6355, and the \CaII\, NIR triplet. About two weeks before the $B$-band maximum, the absorption line near 4300\,\Angst\ could be due to blended \FeII\,$\lambda$4404 and \MgII\,$\lambda$4481, while the broad absorption near 4800\,\Angst\ could be a blend of \FeIII\,$\lambda$5129, \FeII\,$\lambda\lambda\lambda$4924, 5018, 5169, and \SiII\,$\lambda$5051. The distinct absorption features near 3700\,\Angst\ and 7800\,\Angst\ can be identified as \CaII\, H\&K and the \CaII\, NIR triplet, respectively.

After about one week before maximum light, the ``W''-shaped \SII\ absorption features near 5400\,\Angst\ and \SiII\,$\lambda$5972 near 5800\,\Angst\ begin to emerge in the spectra. The minimum of the \SiII\,$\lambda$6355 absorption line shifted redward gradually with the decreasing photospheric velocity, while the IGEs and sulfur gradually gain strength. At around the $B$-band maximum, the spectra are still dominated by the ``W''-shaped \SII\ absorption features and distinct absorption lines of \CaII\, H\&K, \SiII\,$\lambda$6355, and the \CaII\, NIR triplet. The blended absorption lines of \FeIII\,$\lambda$5129, \FeII\,$\lambda\lambda\lambda$4924, 5018, 5169, and \SiII\,$\lambda$5051, as well as the blended absorption lines of \FeII\,$\lambda$4404 and \MgII\,$\lambda$4481, are also notable. At about ten days after maximum light, the ``W''-shaped \SII\ absorption line becomes very weak, while the features of \SiII\,$\lambda$6355 and the 
\CaII\, NIR triplet still remain prominent. At about one month after maximum, the \CaII\, NIR triplet and \CaII\, H\&K absorption lines are still the dominant spectral features. By the time when the SN enters the early nebular phase, features of IGEs start to emerge in the spectra.

\begin{figure*}[ht]
\centering
\includegraphics[angle=0,width=172mm]{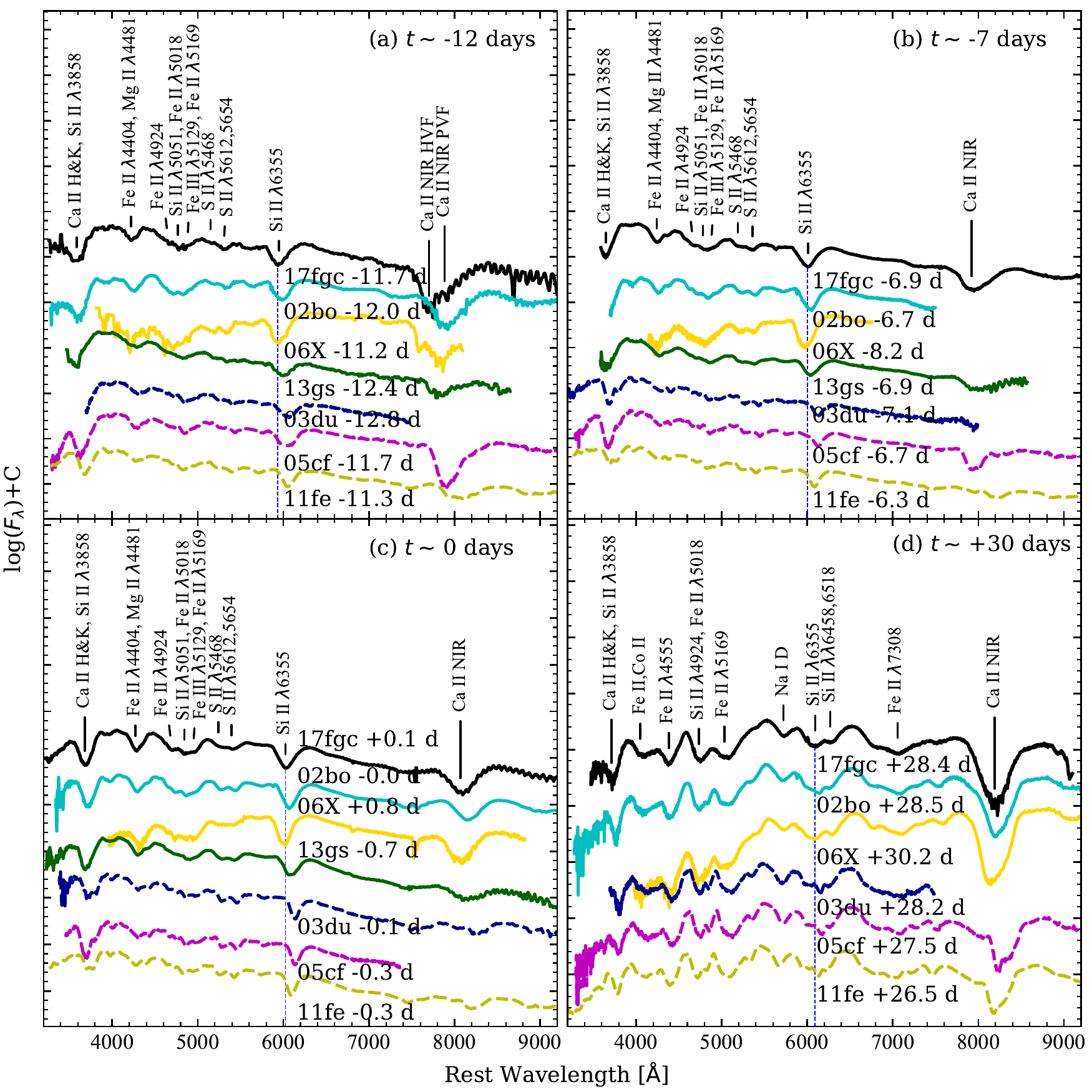}
\caption{Spectra of SN 2017fgc at $t \approx -12$, $-7$, +0, and +30\,d relative to $B$-band maximum, compared with spectra of SNe~2002bo \citep{Benetti2004}, 2003du \citep{Stanishev2007469}, 2005cf \citep{wangxiaofeng2009ApJ}, 2006X \citep{wangxiaofeng2008ApJ}, 2011fe \citep{ZhangKC2016}, and 2013gs \citep{zhangtianmeng2019ApJ} at comparable phases. The spectra of HV SNe Ia are represented with solid lines while those of the NV ones are denoted with dashed lines. All of the above spectra have been corrected for reddening and redshift of the host galaxy.  \label{fspeccomp}}
\end{figure*}

The spectra of SN 2017fgc at four selected epochs are compared with those of several well-observed SNe~Ia having similar \mb, as illustrated in Figure \ref{fspeccomp}. Panel (a) shows the comparison of the $t \approx -12$\,d spectra, including SNe~2017fgc (this paper), 2002bo, 2003du, 2005cf, 
2006X, 2011fe, and 2013gs. 
One can see that both SN 2017fgc and the comparison SNe~Ia have strong absorption features due to \CaII\, H\&K and \SiII~$\lambda$6355. 
The three HV objects (with pEW being 211.1$\pm$19.2\,\Angst\ for SN~2002bo, 211.6$\pm$26.0\,\Angst\ for SN~2006X, and 188.7$\pm$15.6\,\Angst\ for SN~2017fgc, respectively) are found to have systematically stronger $\sim\,4800$\,\Angst\ absorption than three NV ones (with pEW being 86.8$\pm$10.1\,\Angst\ for SN~2003du, 146.8$\pm$3.3\,\Angst\ for SN~2005cf, and 122.2$\pm$4.0\,\Angst\ for SN~2011fe, respectively) at similar phases. While the pEW is measured to be 142.5$\pm$24.4\,\Angst\ for SN 2013gs (HV), which is comparable to that measured for SN 2005cf (NV).
Previous studies suggest that the absorption near 4800\,\Angst\ could be due to a blend of \FeIII\,$\lambda$5129, \FeII\,$\lambda\lambda\lambda$4924, 5018, 5169, and \SiII\,$\lambda$5051 lines.



Figure \ref{fspeccomp}(b) shows the comparison at $t \approx 1$\,week before maximum light. The absorption-line strengths of IMEs are enhanced in the spectra of SN 2017fgc and the comparison SNe~Ia. At this phase, the ``W''-shaped \SII\ absorption features near 5400\,\Angst\ start to emerge in spectra of all SNe~Ia in our sample. However, we notice that the \FeII\ and \MgII\ blended feature near 4500\,\Angst\ is stronger in SNe~2002bo, 2006X, 2013gs, and 2017fgc relative to the three objects in the NV~SN~Ia subclass. 

\begin{figure}[ht]
\centering
\includegraphics[angle=0,width=86mm]{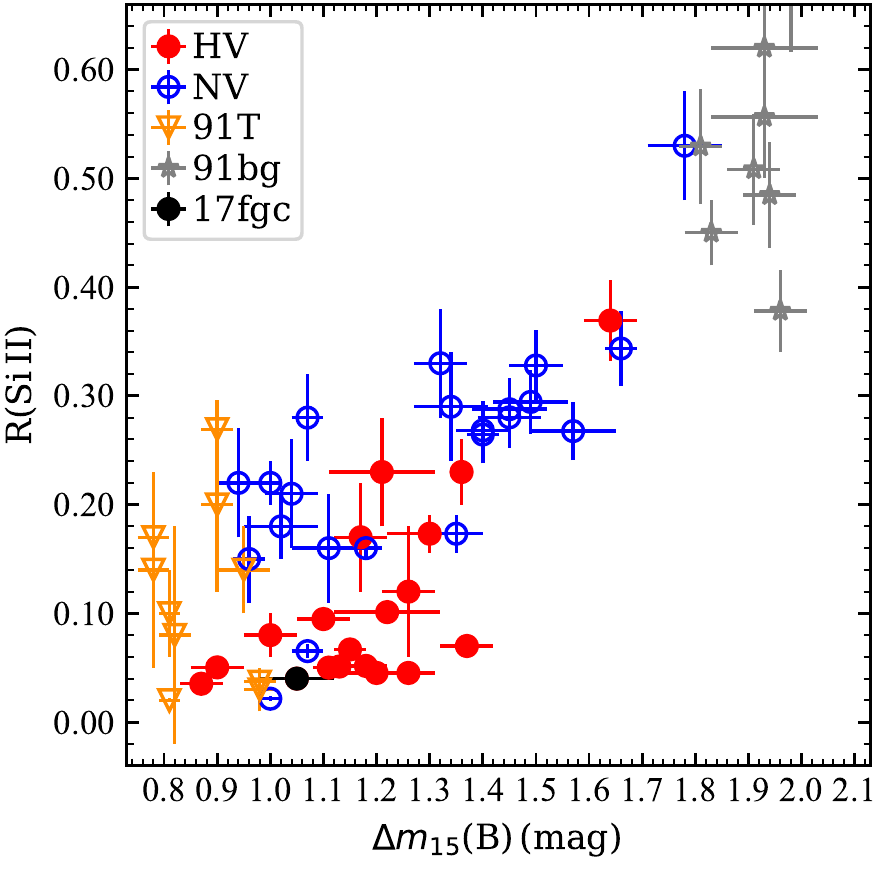}
\caption{Line ratio $R$(\SiII) of several well studied SNe Ia versus luminosity indicator \mb. Some data are taken from \citet{Benetti2004} and \citet{Blondin2012}, while others are taken from \citet{Stanishev2007469} for SN 2003du, \citet{wangxiaofeng2009ApJ} for SN 2005cf, \citet{Zeng2021909} for SN 2017hpa, \citet{liwenxiong2019a} for SN 2018oh, \citet{wangxiaofeng2008ApJ} for SN 2006X, \citet{ZhangTM2010} for SN 2007gi, \citet{zhangtianmeng2019ApJ} for SN 2013gs, and \citet{Kawabata2020893} for SN 2019ein, respectively. The $R$(\SiII) ratios for SNe 1998bp, 1998es, 1999gp, 2001eh, and 2011fe are calculated using the code {\tt respext} with the spectra near maximum light \citep{Stahl2020Spec}. The HV SNe Ia are represented by red dots while the NV ones are shown by blue open circles. The peculiar object SNe 1991bg and 1991T are plotted by red and blue dots, respectively, while SN 2017fgc is overplotted by black dot. \label{fgcRSiII}}
\end{figure}

The spectra near maximum light are displayed in Figure \ref{fspeccomp}(c). 
At this phase, the blended features near 4800\,\Angst\ of SN 2017fgc (with pEW being 227.9$\pm$3.5\,\Angst) are comparable to those of HV objects (with pEW being 204.2$\pm$14.1\,\Angst\ for SN~2002bo, 213.4$\pm$6.4\,\Angst\ for SN~2006X, and 156.6$\pm$9.3\,\Angst\ for SN~2013gs), but they are obviously stronger than those of the NV counterparts (with pEW being 121.1$\pm$13.6\,\Angst\ for SN~2003du, 130.1$\pm$3.2\,\Angst\ for SN~2005cf, and 125.1$\pm$3.2\,\Angst\ for SN~2011fe). The similar situation is found for the blended features near 4300\,\Angst, with pEW being 106.5$\pm$9.6\,\Angst\ for SN~2002bo, 102.2$\pm$4.0\,\Angst\ for SN~2006X, 100.6$\pm$6.1\,\Angst\ for SN~2013gs, 110.7$\pm$2.4\,\Angst\ for SN~2017fgc, and 86.4$\pm$10.0\,\Angst\ for SN~2003du, 92.6$\pm$2.4\,\Angst\ for SN~2005cf, and 85.0$\pm$2.3\,\Angst\ for SN~2011fe.
The velocity measured for SN 2017fgc from the absorption-line minimum of \SiII~$\lambda$6355 at maximum light is \VSiII, which is about 2000 \kms\ faster than that of SN~2002bo (i.e., $\sim$ 13,200 \kms). $R$(\SiII), defined as the line-strength ratio of \SiII~$\lambda$5972 to \SiII~$\lambda$6355 \citep{Nugent1995}, is measured to be \RSiII\ for SN 2017fgc, noticeably smaller than that of the NV SN~2005cf ($R$(\SiII) $= 0.28 \pm 0.04$; \citealp{wangxiaofeng2009ApJ}), perhaps suggesting a relatively higher photospheric temperature of SN 2017fgc around maximum light. 
In Figure \ref{fgcRSiII}, the line ratio \RSiSi\ of a few representative SNe Ia of different subclasses are plotted versus their $B$-band magnitude decline \mb. As expected, the fast decliners tend to have larger \RSiSi\ ratio and vice versa. Compared to the NV SNe Ia with similar decline rates, the HV ones have systematically smaller values of \RSiSi, which provides additional evidence for different properties between these two subclasses. A systematic lower \RSiSi\ ratio indicates that the HV SNe Ia might have higher photospheric temperature and experienced more complete burning or they suffered from interaction of ejecta with CSM or companion stars \citep{Zhao2015220,Zhao2016826}.

Figure \ref{fspeccomp}(d) shows the spectral evolution at $t \approx 30$\,d after maximum light. The absorption profiles of SN 2017fgc and the comaprison SNe~Ia are well developed and tend to have uniform morphologies. 
Form the $t\sim$\,1\,month spectra, the pseudo-equivalent width (pEW) of \CaII\ NIR triplet of SNe 2002bo, 2006X, 2017fgc, 2005cf and 2011fe are measured as 479.0$\,\pm\,$17.5\,\Angst, 475.6$\,\pm\,$9.0\,\Angst, 492.2$\,\pm\,$13.2\,\Angst, 73.9$\,\pm\,$2.9\,\Angst, 89.5$\,\pm\,$0.5\,\Angst, respectively.
We see that the \CaII\ NIR absorption lines in the spectra of HV SNe~2002bo, 2006X, and 2017fgc are much stronger than those of NV SNe 2005cf and 2011fe. 
By $t \approx $1\,month, the \FeII\ features gain strength and gradually dominate the wavelength region 4700--5000\,\Angst.

\begin{figure*}[ht]
\centering
\includegraphics[angle=0,width=172mm]{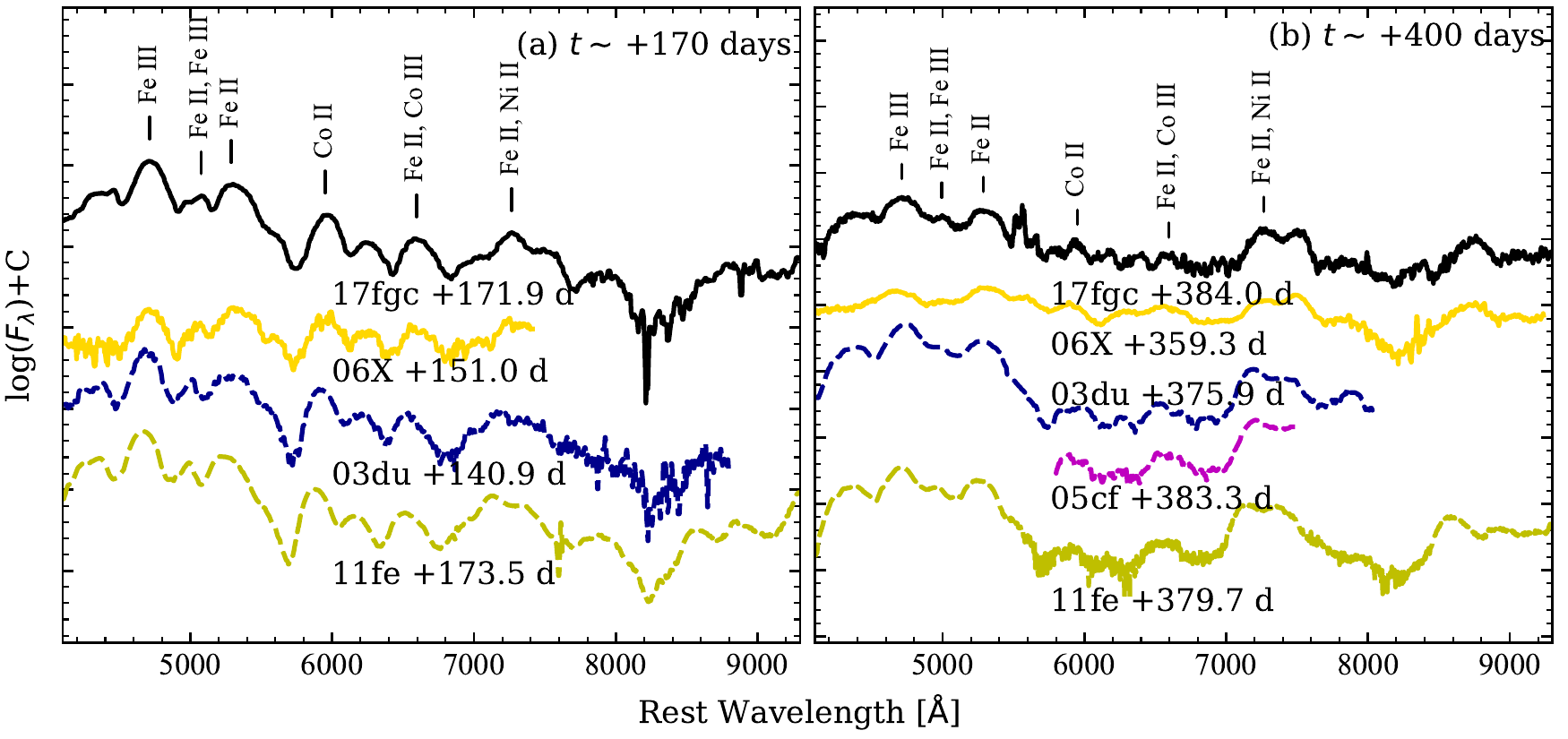}
\caption{Nebular-phase spectra of SN 201fgc at $t \approx +170$\,d and +384\,d relative to $B$-band maximum, compared with spectra of SNe~2003du \citep{Stanishev2007469}, 2005cf \citep{wangxiaofeng2009ApJ}, 2006X \citep{wangxiaofeng2008ApJ}, 2011fe \citep{ZhangKC2016}, and 2013gs \citep{zhangtianmeng2019ApJ} at comparable phases. The HV SNe Ia are represented with solid lines while the NV ones are dashed lines. All of the spectra have been corrected for reddening and redshift of the host galaxy.  \label{fspeccomp2}}
\end{figure*}

Nebular spectra at $t \approx 170$\,d and $t \approx 400$\,d after maximum light are shown in Figures \ref{fspeccomp2}(a) and \ref{fspeccomp2}(b), 
respectively. The spectra of SN 2017fgc are well developed at such late phases and characterized by the forbidden lines of singly and doubly ionized IGEs, such as the [\FeII] and [\FeIII] features at $\sim 4700$, $\sim 5000$, $\sim 6500$, and $\sim 7000$\,\Angst, as well as [\CoII] and [\CoIII] at  $\sim 5800$ and  $\sim 6500$\,\Angst. The same lines are commonly seen in the comparison SNe~Ia at similar phases.
We notice that the emission profile of [\FeII]/[\FeIII] at $\sim$5000\,\Angst\ is stronger in the NV SNe Ia (i.e., with the pEW being measured as 63.8$\,\pm\,$2.3\,\Angst\ and 62.8$\,\pm\,$1.6\,\Angst\ for SN 2003du and SN 2011fe, respectively) than in HV ones (i.e., with the pEW being measured as 24.8$\,\pm\,$7.3\,\Angst\ and 26.0$\,\pm\,$1.9\,\Angst\ for SN 2017fgc and SN 2006X, respectively).
From the $t \approx +384$\,d spectrum, we measured the velocity shift of [\FeII]\,$\lambda$7155 and [\NiII]\,$\lambda$7378 as $+2220 \pm 260$\,\kms, and that measured from the $t \approx +389$\,d spectrum is $+1640 \pm 580$\,\kms; these are consistent with the trend that HV~SNe~Ia tend to have redshifted \FeII/\NiII\ velocity in the nebular phase \citep{Maeda2010Natur,Maguire2018477}.


\subsection{Ejecta Velocity}
\begin{figure}[ht]
\centering
\includegraphics[angle=0,width=86mm]{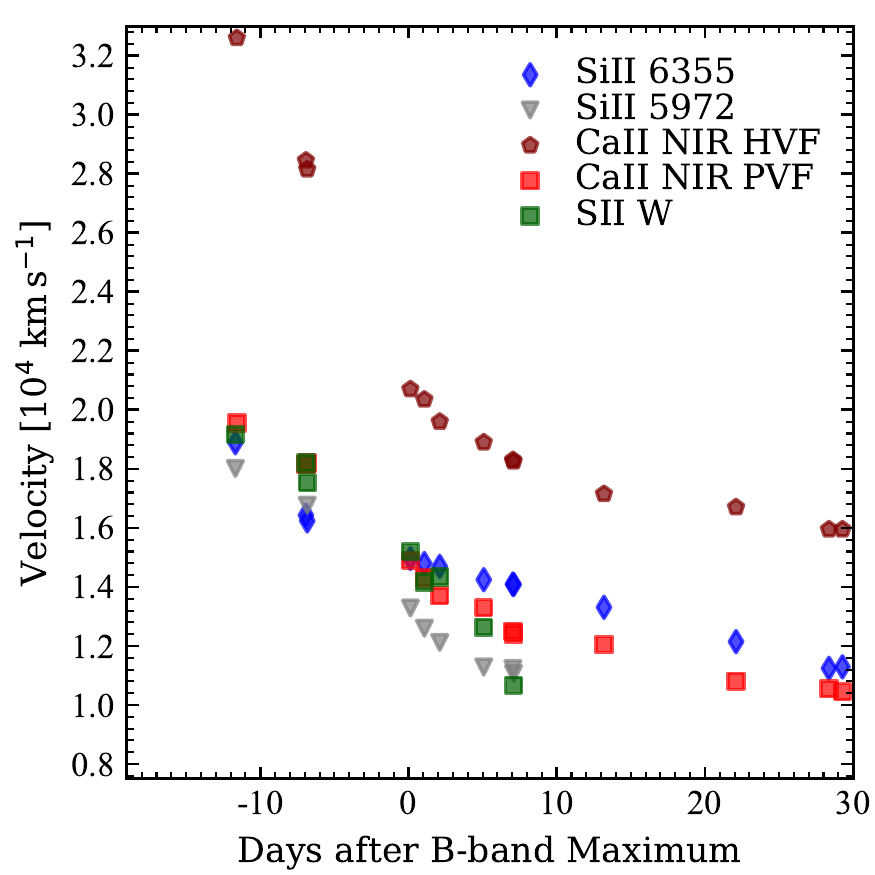}
\caption{Velocity evolution of different elements measured from spectra of SN 2017fgc. Note that for \CaII, both the detached high-velocity feature (HVF) and photospheric-velocity feature (PVF) are shown for comparison. 
\label{felementV}}
\end{figure}

The methods described by \citet{Zhao2015220,Zhao2016826} were utilized to measure the ejecta velocity from the absorption lines, such as \SII\,$\lambda\lambda$5460, 5640, \SiII\,$\lambda$5972, \SiII\,$\lambda$6355, and the \CaII\,NIR triplet, are shown in Figure \ref{felementV}. The photospheric velocity measured from \SiII\,$\lambda$6355 at $t \approx -11.7$\,d is $\sim 18,900$\,\kms, comparable to that of the \SiII\,$\lambda$5972 ($\sim 18,000$\,\kms) and \SII\ lines ($\sim 19,000$\,\kms), but much slower than the high-velocity-feature (HVF) velocity inferred from the \CaII\ NIR triplet ($\sim 32,000$\,\kms). At the time of $B$-band maximum, the velocity of \SiII\,$\lambda$6355 is measured as \VSiII, significantly larger than the typical value (i.e, $\sim 11,800$\,\kms) of normal velocity (NV) SNe~Ia. We thus put SN 2017fgc into the HV subclass according to the classification criteria proposed by \citet{wangxiaofeng2009ApJL}.

\begin{figure}[ht]
\centering
\includegraphics[angle=0,width=86mm]{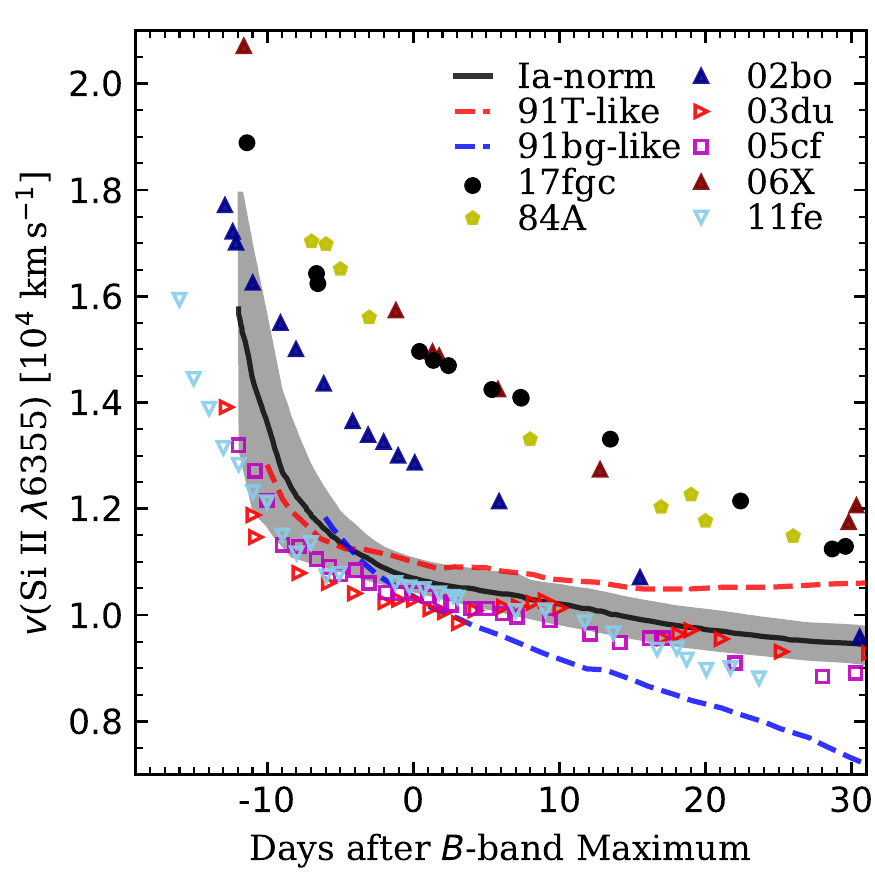}
\caption{Evolution of the expansion velocity of SN 2017fgc as derived from the minimum of the \SiII\,$\lambda$6355 absorption line, compared with those of HV sample of SNe Ia, including 1984A, 2002bo, 2006X, and NV sample including SN 2003du, SN 2005cf, SN 2011fe. The average velocity curves obtained for SN 1991T-like and SN 1991bg-like SNe are overplotted with red and blue dashed lines, respectively. The normal subclass of SNe~Ia is plotted with a black solid line. The shaded region represents the 1$\sigma$ uncertainty for the mean velocity curve of NV SNe~Ia. The values of SN 1984A are taken from \citet{Barbon1989}, SN 2002bo from \citet{Benetti2004}, SN 2006X from \citet{wangxiaofeng2008ApJ}, SN 2005cf from \citet{wangxiaofeng2009ApJ}, and SN 2011fe from \citet{ZhangKC2016}. The region of NV SNe~Ia is extracted from \citet{wangxiaofeng2009ApJL}. 
\label{fSiIIV}}
\end{figure}

Applying a linear fitting to the velocities of \SiII\,$\lambda$6355 measured during the period $t \approx 0$\,d to $t \approx +10$\,d, we derive the velocity gradient as \VSiIIDot, suggesting that SN 2017fgc belongs to the HVG subclass. Figure \ref{felementV} shows the velocity evolution of some IMEs, including \SII, \SiII, and \CaII. The velocities of \SiII\,$\lambda$5972 and the \CaII\,NIR triplet photospheric component show similar 
evolution, while \SiII\,$\lambda$6355 exhibits slightly higher velocities. The HVF of the \CaII\,NIR triplet has the highest velocity. The photospheric velocity evolution of SN 2017fgc, as derived from the minimum of \SiII\,$\lambda$6355 absorption, is shown in Figure \ref{fSiIIV}, together with that of the comparison SNe~Ia, including SNe~2002bo, 2005cf, 2006X, and 2011fe. As can be seen, SN 2017fgc exhibits a velocity evolution similar to that of SN 2006X. The basic photometric and spectroscopic parameters of SN 2017fgc are listed in Table \ref{tabpars}.


\section{Discussion}
\subsection{Velocity and Velocity Gradient}

\begin{figure*}[ht]
\centering
\includegraphics[angle=0,width=172mm]{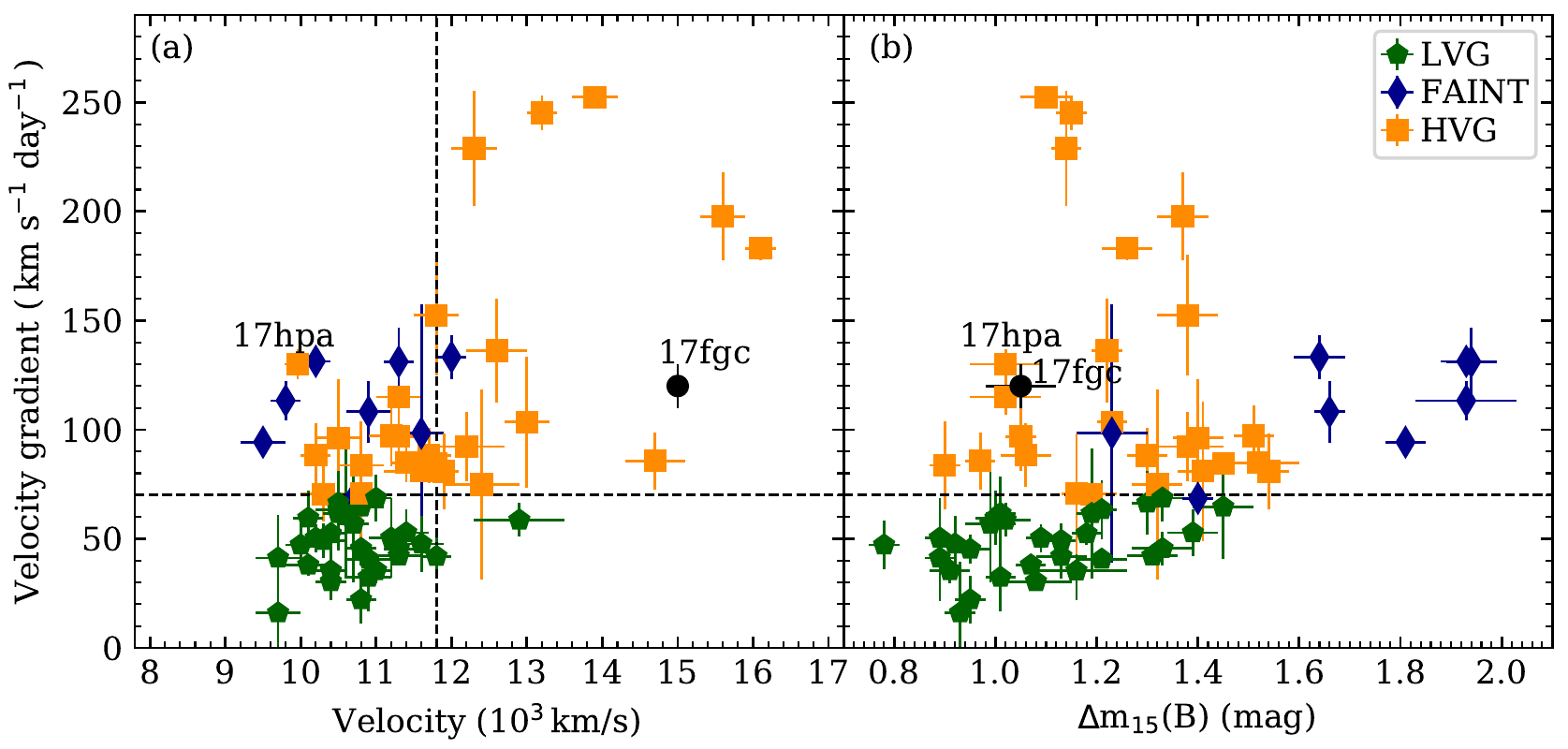}
\caption{The spectroscopic subclassification of SN 2017fgc (as marked with a black dot) based on the scheme of \citet{Benetti2005}. (a) The scatter plot of the \SiII\ velocity versus the velocity gradient. The velocity gradients of SN 2005cf and SN 2018oh are taken from \citet{wangxiaofeng2009ApJ} and \citet{liwenxiong2019a}, respectively, and those of the other objects are from \citet{Benetti2005} and \citet{Folatelli2013773}. The velocities are taken from \citet{Silverman2012c} and \citet{wangxiaofeng2019ApJ}. The horizontal dashed line in the right pannel marks the boundary between HVG and LVG, which is 70\,km\,s$^{-1}$\,d$^{-1}$ as defined by \citet{Benetti2005}, while the vertical dashed line represents the boundary 
between the HV and NV~SNe~Ia as defined by \citet{wangxiaofeng2009ApJ}. (b) \mb\ is plotted versus the velocity gradient measured from the \SiII\,$\lambda$6355 absorption line in the near-maximum-light spectra. \label{HVG1}}
\end{figure*}

The large velocity ($\sim$ \VSiII) and velocity gradient ($\sim$ \VSiIIDot) clearly put SN 2017fgc into the subclasses of both HV and HVG. In general, most HV~SNe~Ia belong to the HVG subclasses, while the NV~SNe~Ia correspond to the LVG ones \citep{wangxiaofeng2009ApJL,Silverman2012c}, as shown in Figure \ref{HVG1}(a). However, there are also outliers such as SN 2017hpa, which has a normal velocity around maximum light ($\sim 9550$\,\kms) but a large velocity gradient ($\sim 130$\,\kms) \citep{Zeng2021909}. This indicates that HVG~SNe~Ia could have multiple physical origins. This is also demonstrated by the correlation between velocity gradient and \mb\ \citep{Phillips1993}, as shown in Figure \ref{HVG1}(b), where no obvious correlation exists between velocity gradient and the luminosity indicator \mb\ while only the LVG sample possibly shows a weak correlation between these two observables.

Previous studies have shown that the difference in velocity gradient may be related to the different nature of the explosions or the mixture of SN~Ia ejecta (e.g., \citealp{Benetti2004, Sahu2013}). Different viewing angles in an off-center ignition during the explosion of SNe~Ia will lead to variations in the observed velocity gradient \citep{Maeda2010Natur}. However, this explanation has difficulty explaining the observed fact that HVG (HV) SNe~Ia preferentially occur in the inner region of the galaxy \citep{wangxiaofeng2013Sci}. \citet{Woosley2009} and \citet{Blondin2012} found that varying the criterion of the deflagration-to-detonation transition (DDT) will cause varying velocity gradients in a wide range. The HVG subclass may result from adequate mixing of heavy elements in the SN ejecta, while the counterparts with low velocity gradient may suffer less mixing in the ejecta \citep{Blondin2012, Sahu2013}. As suggested by \citet{Zeng2021909}, the large velocity gradient seen in SN 2017hpa could be caused by the effective mixing of heavy elements in the SN ejecta.


\subsection{Light-Curve Features}

\begin{figure*}[ht]
\centering
\includegraphics[angle=0,width=172mm]{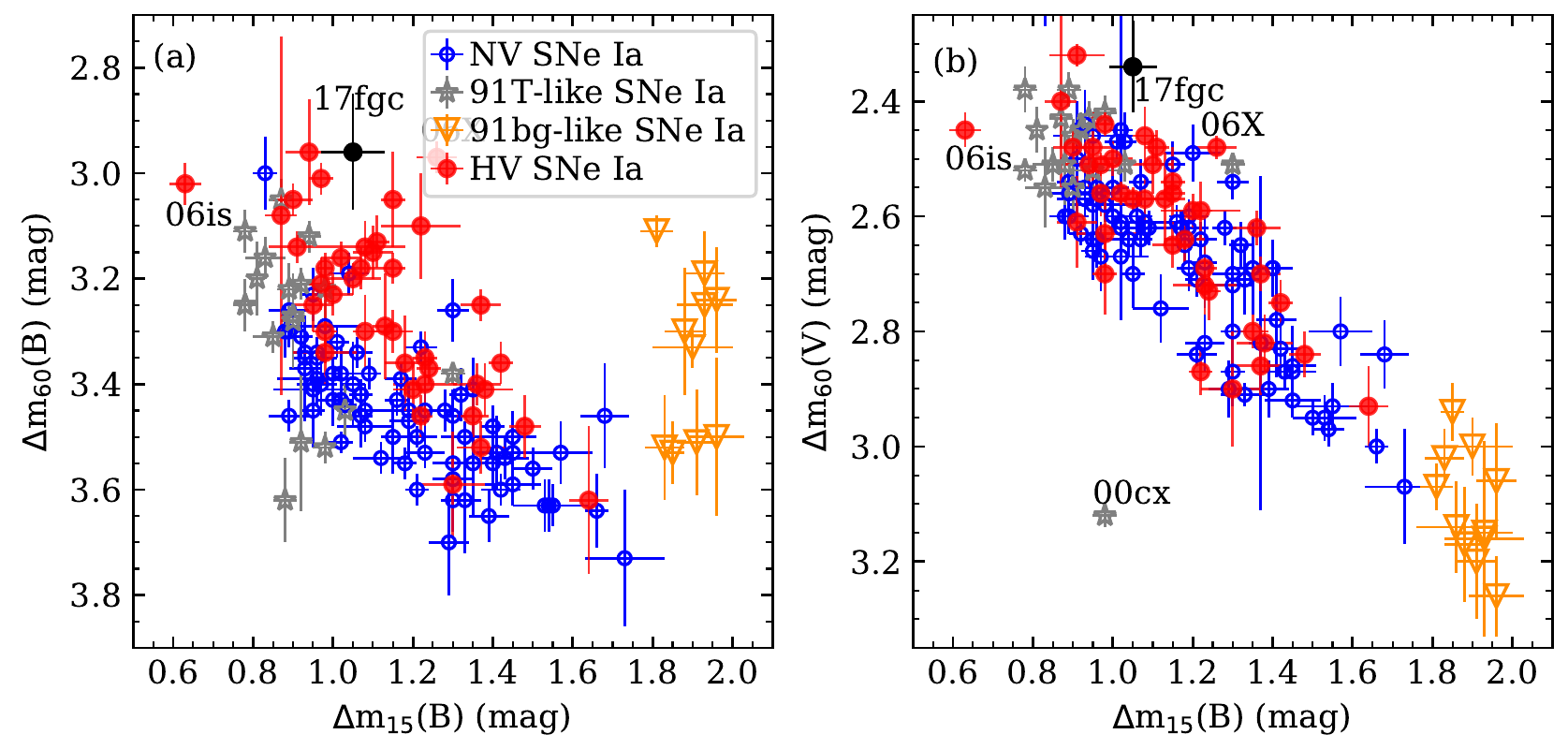}
\caption{Tail brightness of SNe~Ia, measured as the magnitude decline at $t \approx 60$\,d from the peaks of the (a) $B$ and (b) $V$ light curves, 
versus the luminosity indicator \mb\ that is measured as the magnitude decline within the first 15\,d after $B$-band maximum \citep{Phillips1993}. 
The data are taken from \citet{wangxiaofeng2019ApJ}, while SN 2017fgc is overplotted as a black dot. \label{HVG2}}
\end{figure*}

During their study of the typical HV object SN 2006X, \citet{Wang2008677} 
noticed that it exhibited relatively flat tail evolution starting around +40\,d after maximum light. A recent statistical study based on a large sample indicates that the excess blue flux is common for the HV group of SNe~Ia \citep{wangxiaofeng2019ApJ,LiWX2021906}. For SN 2017fgc, the decline from peak brightness measured at $t \approx 60$\,d is 2.96\,mag in $B$ and 2.34\,mag in $V$. The corresponding values are much smaller than in NV SNe~Ia with similar \mb, but are consistent with the behavior of HV~SNe~Ia. Figure \ref{HVG2} shows the results measured for SN 2017fgc and the comparison sample from \citet{wangxiaofeng2019ApJ}. For the NV sample 
of SNe~Ia, more-luminous objects tend to have brighter tails and slower decline rates, while this tendency shows large scatter in the $B$ band owing to the abnormally bright tails exhibited by the HV subsamples \citep{wangxiaofeng2019ApJ}. This is primarily due to the fact that HV~SNe~Ia tend to show excess emission in the early nebular phase, which can be caused 
by light scattering by the surrounding CSM \citep{wangxiaofeng2019ApJ}. This scenario is favored by the detections of nearby light echoes around SN 2006X \citep{Wang2008677} and SN 2014J \citep{Yang2018854}.

\begin{figure*}[ht]
\centering
\includegraphics[angle=0,width=172mm]{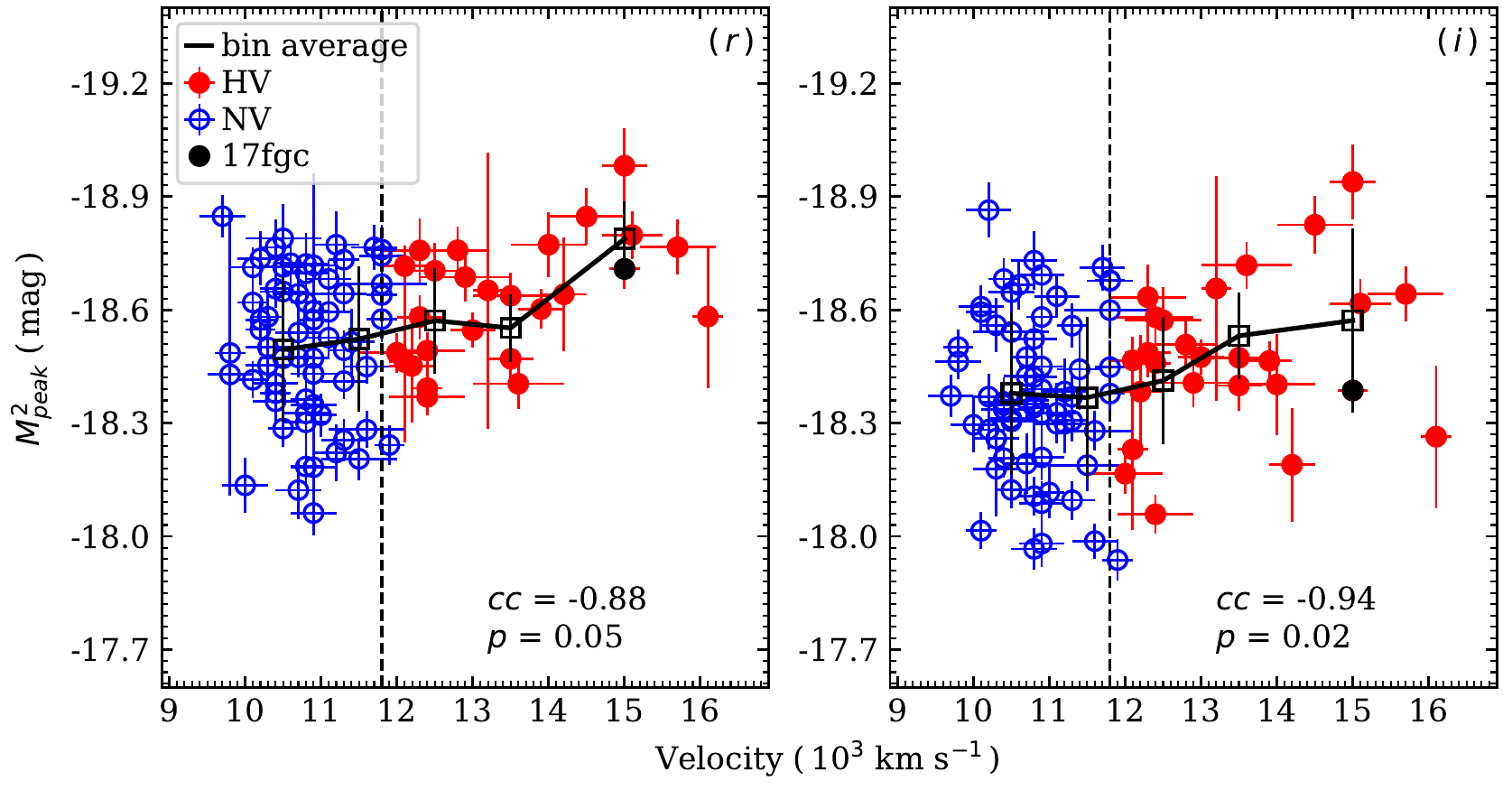}
\caption{{\it Left:} The absolute secondary peak magnitude in $r$ band is plotted as a function of the \SiII\ velocity of SNe~Ia. The velocities are taken from \citet{wangxiaofeng2019ApJ}. The black open square with black line represents the binned average, and $cc$ and $p$ are the correlation coefficient and p-value from pearson statistics \textbf{(with significance level $\sim$2$\sigma$),} respectively. {\it Right:} Same, but in the $i$ band. The HV SNe Ia are represented with red dots while the NV ones are in blue open circles. The vertical dashed lines represent the boundary between the HV and NV~SNe~Ia as defined by \citet{wangxiaofeng2009ApJ}. \label{fgcpVMpeak}}
\end{figure*}

Besides the flatter bluer-band light-curve evolution in the early nebular 
phase, HV~SNe~Ia seem to share another light-curve feature: a stronger secondary shoulder or maximum in the $R/r$ and $I/i$ bands. This seems to apply for SN 2017fgc. Following \citet{Stahl2020Spec}, the Gaussian-process light-curve fitting implemented in SNooPy2 is employed to obtain the secondary peak magnitudes of SN~Ia light curves in the $R/r$ and $I/i$ bands. The resulting absolute $RrIi$-band secondary peak magnitudes of 92 SNe~Ia (including 26 HV and 66 NV~SNe~Ia) are plotted as a function of \SiII\ velocity in Figure \ref{fgcpVMpeak}. 
The black line with square symbols represent the binned average (i.e., a bin of 1,000\,\kms is used to calculate the average value). The pearson coefficient is estimated as -0.94 and -0.88 for the binned $r$-band and $i$-band data, respectively. The corresponding p-values are 0.02 and 0.04, respectively, suggesting that the shoulder/secondary peak features could be correlated with the \SiII\ velocity at a confidence level of $\sim$2$\sigma$.



The NIR light curves of SNe~Ia have been occasionally studied theoretically. Several delayed-detonation models have been applied by \citet{Hoflich1995443} to reconstruct the $I$-band and NIR light curves. 
They described the double-peaked behavior as a temperature effect, in which the drop in temperature of the ejecta is compensated by the expansion of the photosphere. \citet{Pinto2000530} explained the secondary maximum in the NIR bands as the release of trapped radiative energy owing to the decrease in the flux mean opacity. \citet{Kasen2006649} found that the NIR secondary maximum of SNe~Ia could be due to the ionization transition of IGEs in the ejecta from doubly to singly ionized, which in turn leads to a weakening of \FeIII\ and \CoIII\ lines and a strengthening of \FeII\ and \CoII\ lines (see also \citealp{Jack2012538,Dessart2014441,Blondin2015448}). They suggested that the properties of the secondary maximum are related to \mb\ (and hence the peak luminosity), being more prominent in the brighter SNe~Ia. Moreover, a small fraction of stable IGEs may be produced depending on the metallicity of the progenitor WD during the burning to nuclear statistical equilibrium \citep{Kasen2006649,Kasen2007656}. 
The peak magnitude of the secondary maximum increases with the growing of the stable iron core, while the increase in metallicity of the progenitor increases the size of the iron core. 
In comparison NV SNe Ia, the more pronounced shoulder/secondary peak features of HV SNe Ia seen in $R/r$ and $I/i$ bands might be related to that the latter subclass have more metal-rich progenitor populations \citep{wangxiaofeng2013Sci,Pan2020895}.

\subsection{Absorption of Intermediate-Mass and Iron-Group Elements}

It is common that HV~SNe~Ia also show stronger absorption of \SiII\,$\lambda$6355 as evidenced by the fact that this subclass overlaps greatly with the BL subclass.
According to \citet{Kasen2006649}, the SN ejecta becomes very effective in redistributing the blue/UV photons to longer wavelengths when the iron-rich layers of the SN ejecta cools down to $\sim$7000 Kelvin, which leads to the rebrightening of the NIR light curves. \citet{Jack2012538} also found that the recombination of \FeIII\ to \FeII\ is responsible for the second maximum features in NIR bands. \citet{Silverman2012d} suggested that the pEW of \MgII\ complex and \FeII\ complex are correlated to the SALT2 colour, with the HV SNe being redder and having larger pEWs (see also \citealp{Nordin2011526,Walker2011410}).
To further investigate the discrepancy between HV and NV~SNe~Ia (including 66 NV and 26 HV ones), we examine the pEW of the blended absorption near 4400\,\Angst (including \FeII\,$\lambda$4404 and \MgII\,$\lambda$4481) and that near 4900\,\Angst (including \FeIII\,$\lambda$5129, \FeII\,$\lambda\lambda\lambda$4924, 5018, 5169, and \SiII\,$\lambda$5051), respectively dubbed as pEW(Mg\,II) and pEW(Fe\,II). 
The samples are the same as those used by \citet{wangxiaofeng2019ApJ}. The code {\tt respext}\footnote{https://github.com/benstahl92/respext} \citep{Stahl2020Spec} is employed to estimate the pEWs of \FeII\,$\lambda$4404/\MgII\,$\lambda$4481 and \FeII/\FeIII\ blends from the spectra around maximum light, and linear fitting is utilized to infer the pEWs at $B$-band maximum. 

\begin{figure*}[ht]
\centering
~
\includegraphics[angle=0,width=84.5mm]{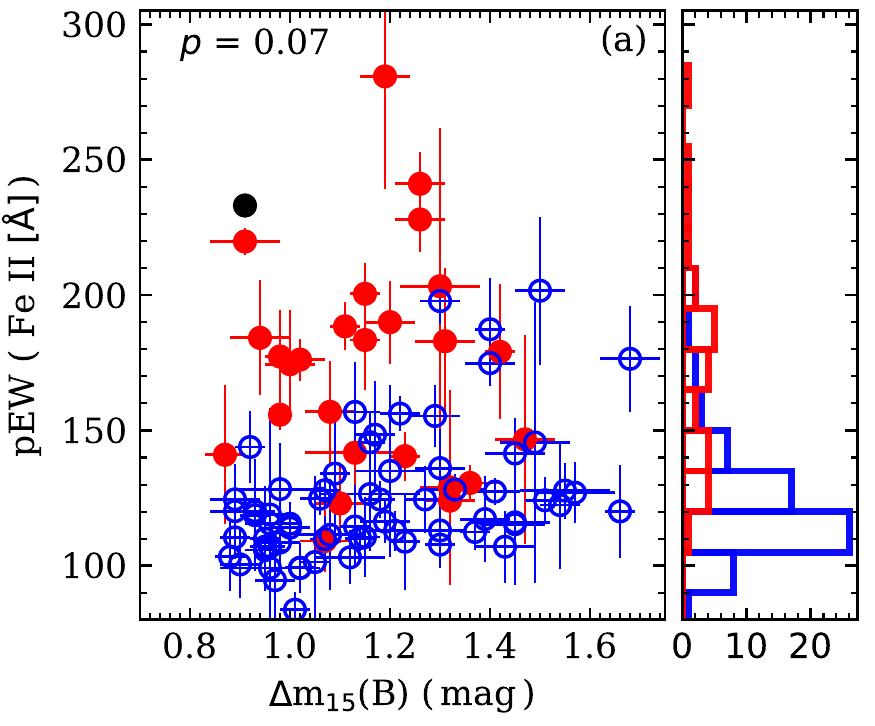}
~
\includegraphics[angle=0,width=84.5mm]{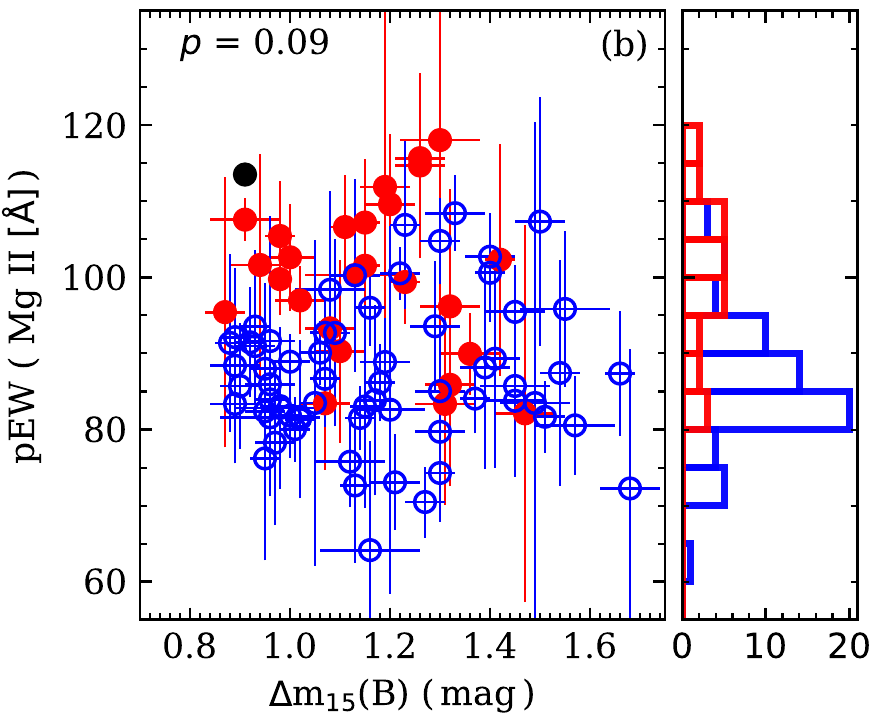}
\\
\includegraphics[angle=0,width=172mm]{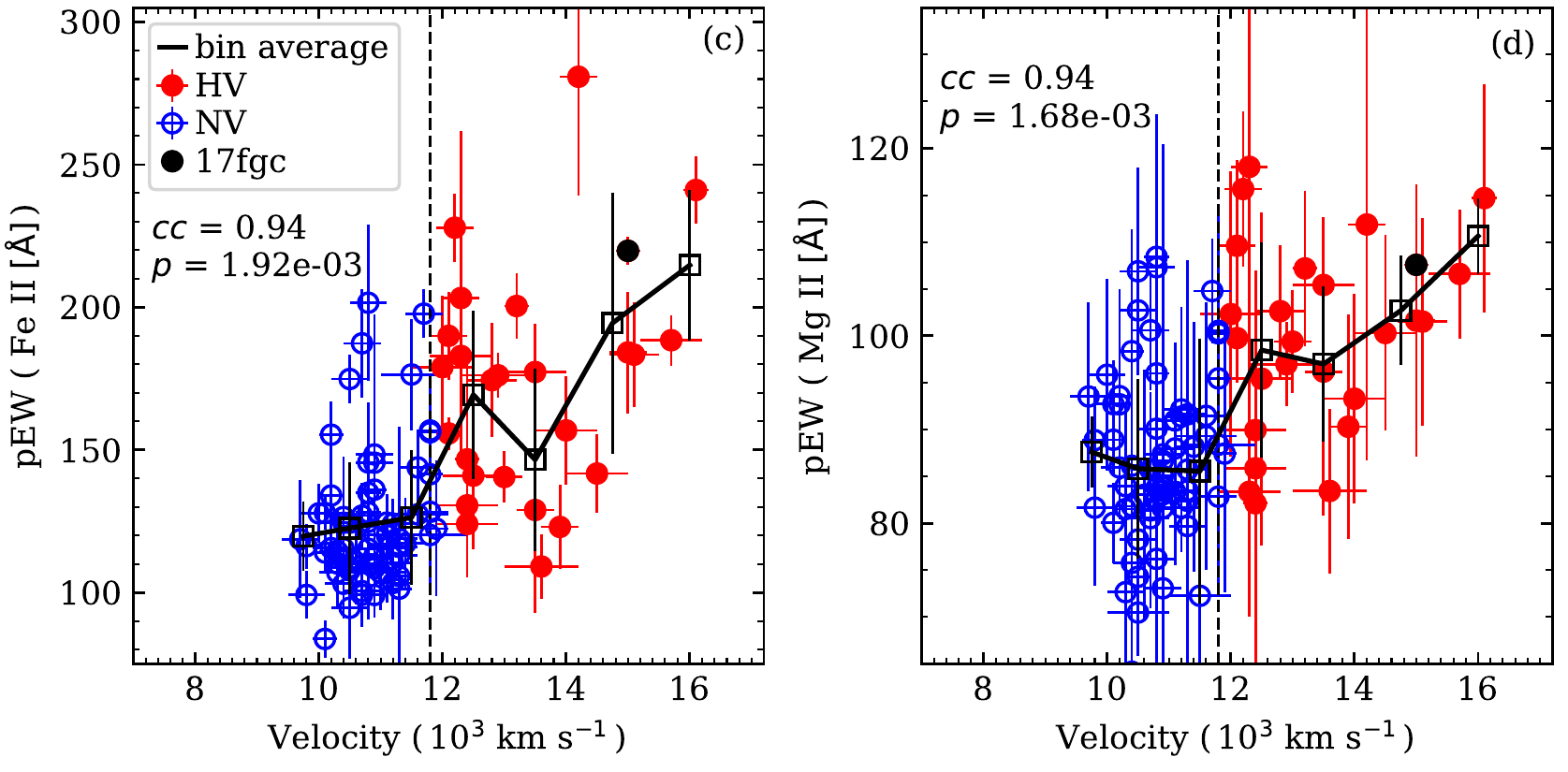}
\caption{Panels (a) and (b): The pseudo-equivalent widths (pEWs) of blended absorptions of \FeIII\,$\lambda$5129, \FeII\,$\lambda\lambda\lambda$4924, 5018, 5169, and \SiII\,$\lambda$5051 measured in the near-maximum-light spectra are plotted versus the luminosity indicator \mb\ for 92 SNe~Ia from \citet{wangxiaofeng2019ApJ}. \textbf{The $p$ represents the p-value (with significance level $\sim$2$\sigma$) from the $T$-test statistics.} Panels (c) and (d): The pEWs of the blended absorptions of \FeII\,$\lambda$4404 and \MgII\,$\lambda$4481 are plotted as a function of \SiII\ velocity obtained around $B$-band maximum. The HV~SNe~Ia are represented with red dots while the NV subclass are blue open circles. For comparison, SN 2017fgc is shown with a black dot. The \SiII\ velocities of SNe~Ia are taken from \citet{wangxiaofeng2019ApJ}.
The black open square with black line represents the binned average, and $cc$ and $p$ are the correlation coefficient and p-value from pearson statistics (with significance level $\sim$3$\sigma$), respectively. The vertical dashed lines represent the boundary between the HV and NV~SNe~Ia as defined by \citet{wangxiaofeng2009ApJ}. \label{fgcpEWdmv}}
\end{figure*}

Figure \ref{fgcpEWdmv} (a) and (b) show the correlations between pEW(\MgII)/pEW(\FeII) and \mb. One can see that HV~SNe~Ia have on average larger pEW(\MgII) and pEW(\FeII) relative to NV~SNe~Ia.
The p-values from the T test of the pEW measured for HV and NV SNe Ia are 0.07 and 0.09, corresponding to a significance of $\sim 2\sigma$.
This indicates that HV and NV SNe Ia may have different ejecta properties that might be related to explosion mechanism or progenitor system.
Inspection on panels (c) and (d) of Figure \ref{fgcpEWdmv} reveals that both pEW(\FeII) and pEW(\MgII) have a positive correlation with the \SiII\ velocity. The difference in absorption features of IMEs and IGEs suggests that the outer ejecta of the HV~SNe~Ia may have experienced more complete burning than the NV subclass. This seems to be consistent with the delayed-detonation model, which involves an initial subsonically propagating mode of nuclear burning followed by a supersonically moving detonation front with some time delay \citep{Khokhlov1989239}. Several three-dimensional models based on delayed detonation Chandrasekhar-mass explosions also have been proposed in variants of DDT models \citep{Khokhlov20052006,Bravo2008478,Ropke2007464,Ropke2012750}. The reduced burning densities resulting from energy release in the subsonical propagation mode of nuclear burning leads to extensive burning of the remaining fuel \citep{Seitenzahl2013}. In addition, larger off-center distances of DDT models might lead to more silicon at high velocities \citep{Hoflich2002568,Hoflich200650}. For a given \mb, the asymmetric detonation can aspherically push Si outward and make the \SiII\ features form in the regions at higher velocities \citep{Hoflich200650,WangLF2007315,Yang2018854,Cikota2019490}.


As an alternative, the larger pEWs seen in HV objects might be related to their environment metallicity, which could play a role in affecting the observed properties of SNe~Ia \citep{Dominguez2001557,Timmes2003590,Silverman2012c}.
Higher metallicity for HV~SNe~Ia is favored by the study of their birthplace environments \citep{wangxiaofeng2013Sci,Pan2020895}. 
According to \citet{wangxiaofeng2013Sci}, HV SNe Ia likely have younger massive and metal-rich progenitor system than NV ones. The increase in stellar metallicity of progenitors could partially cause the higher \SiII\ velocity seen in HV SNe Ia. The metal-rich stars could produce stronger outflows and produce more abundant CSM than the metal-pool counterparts \citep{wangxiaofeng2013Sci}, which is consistent with the observations that those showing blueshifted Na ID absorptions are more likely to be HV SNe Ia \citep{Sternberg2011333,Foley2012744,wangxiaofeng2019ApJ}.
The strong absorption lines of \FeII\ and IMEs seen in the spectra of SN 2017fgc could thus originate from the metal-rich progenitor system, which will be further addressed below.

\subsection{Explosion Environment and its Metallicity}\label{sec54}

\begin{figure*}[ht]
\centering
\includegraphics[angle=0,width=172mm]{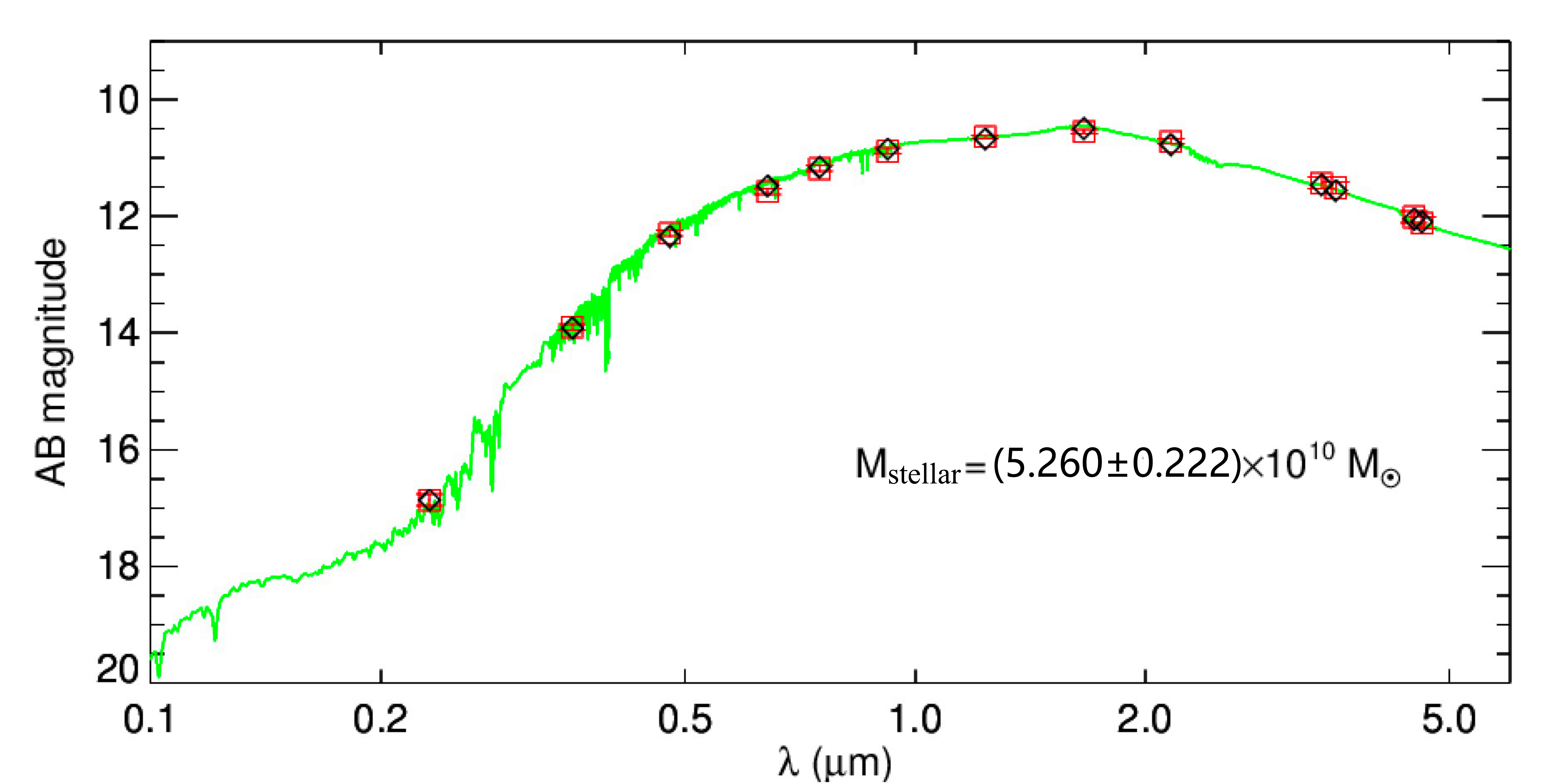}
\caption{The stellar population synthesis fit for the SED of NGC474. The red open squares represent the observed AB magnitudes, while the synthesized magnitudes from the best-fit spectrum template (green curve) are overplotted as black diamonds. \label{fgcHostSED}}
\end{figure*}

In order to study the properties of the host galaxy NGC 474, a total of 13 bands of photometric data ranging from the UV to the IR (including the NUV band from {\it GALEX}, five broad bands from SDSS, three NIR bands from 2MASS, and four NIR and mid-IR bands from {\it Spitzer} and {\it WISE}) from the NASA/IPAC extragalactic database (NED\footnote{https://ned.ipac.caltech.edu/}) have been downloaded to construct its spectral energy distribution (SED). The stellar population synthesis code BC03 \citep{Bruzual2003344} was employed to construct the spectral models with the adopted 
parameter configurations, including the initial mass function (IMF; \citealp{Chabrier2003115}) and the Padova 1994 evolutionary tracks and delayed-exponential star-formation history. 
The adopted SED-ftting method is described in \citet{Wei202121} while a $\chi ^2$ minimization is used to fit the total stellar mass of NGC 474 (see also \citealp{Lin2013769}). The observed SEDs and the best-fit templates of NGC 474 are shown in Figure \ref{fgcHostSED}.
The logarithmic stellar mass of NGC 474 is estimated to be 10.72$\,\pm\,$0.02 by fitting the observed SED with the stellar population synthesis model. Utilizing the {\it Spitzer} observations at 3.6\,$\mu$m, \citet{Alabi2020497} derived a similar logarithmic stellar mass of NGC 474 as $\sim 10.6$.

According to \citet{Wu2019484}, the relation between host galaxy stellar mass and metallicity can be characterized empirically as

\begin{figure}[ht]
\centering
\includegraphics[angle=0,width=86mm]{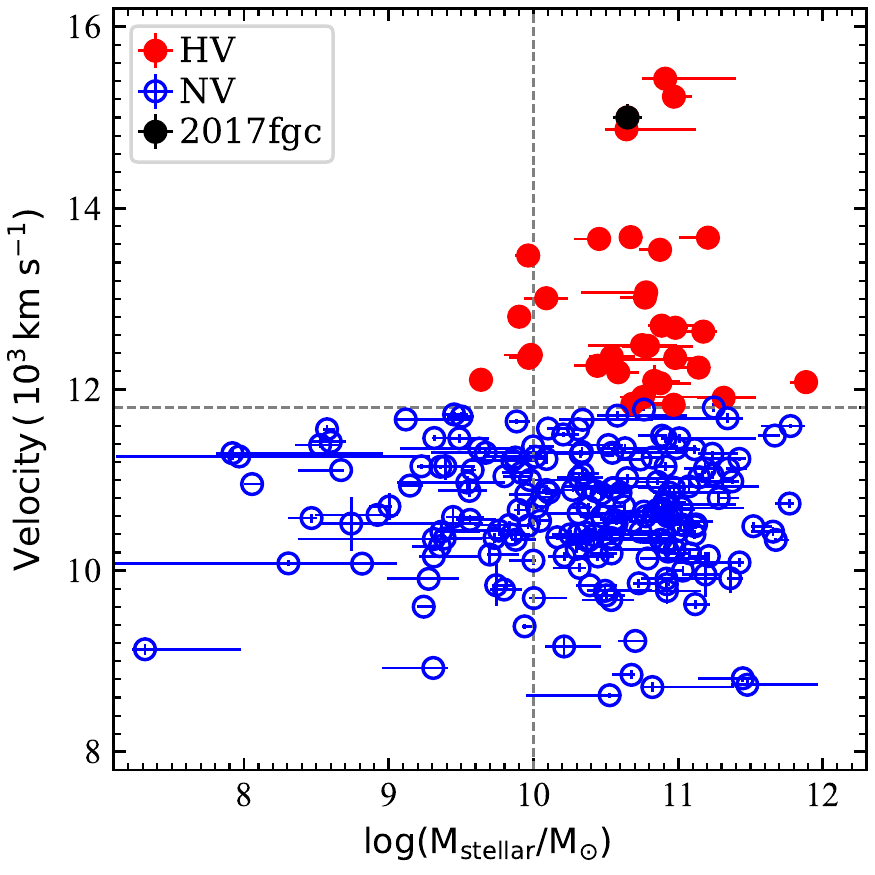}
\caption{Scatter plot of the \SiII\,$\lambda$6355 velocities and the corresponding host-galaxy stellar mass ($M_{\rm stellar}$). The HV~SNe~Ia are shown as red dots, and the NV~SNe~Ia are shown as blue circles. The horizontal and vertical dashed lines represent the criteria used to split the sample in velocity \citep{wangxiaofeng2009ApJL} and $M_{\rm stellar}$ \citep{Pan2020895} space, respectively. The data are taken from \citet{Pan2020895} and SN 2017fgc (this paper) is overplotted as a black dot. \label{fgcHost}}
\end{figure}

\begin{footnotesize}
\begin{equation} 
\label{mzeq}
Z=-1.492+1.847\,{\rm log}(M_{*}/M_{\sun})-0.08026 \big [{\rm log}(M_{*}/M_{\sun} \big ]^2,
\end{equation}
\end{footnotesize}
\noindent
where $Z$ is the oxygen abundance 12 + log(O/H). Using this relation, a supersolar oxygen abundance of $9.08 \pm 0.12$ can be derived for NGC 474. From Figure \ref{fgcHost}, one can see that SN 2017fgc is consistent with the finding that HV~SNe~Ia tend to occur in more massive and metal-rich host galaxies. Also, the spectral survey for NGC 474 by \citet{Alabi2020497} shows that this galaxy has a slightly more metal-rich outer shell compared to the galaxy center. Recently, the detailed study by \citet{Fensch2020644} showed that NGC 474 merged with a young metal-rich galaxy during its evolution.



Although a higher metallicity may help explain the observed properties seen in SN 2017fgc and some HV~SNe~Ia, we notice that SN 2017fgc is located far away ($\sim 18.90 \pm 0.01$\,kpc) from the center of its host galaxy. 
\citet{wangxiaofeng2013Sci} initially found that HV SNe Ia tend to occur in more massive galaxy and explode near the center of the host galaxy, which is later confirmed by the Palomar Transient Factory (PTF) and the Berkeley SN Ia program sample \citep{Pan2015446,Pan2020895}. It seems that SN 2017fgc is an outlier and does not follow this trend of HV SNe Ia. However, closer inspection of the host galaxy NGC 474 reveals that it is a massive lenticular galaxy that experienced a merger $\sim 2$\,Gyr ago \citep{Alabi2020497,Fensch2020644}. SN 2017fgc is located in a gas bridge (see the left panel of Fig. 1) connecting the gas shell and the massive galaxy. We speculate that SN 2017fgc could be ejected from the inner part of the companion galaxy NGC 470 during the merger took place at $\sim$2 Gyr ago, or formed as a result of some cold gas remained in the companion disk.

We attempted to measure the metallicity of the explosion site from both the multicolor photometry and MUSE IFU \citep{Laurent20107739} spectra but 
failed because no spectra near SN 2017fgc are available. 
According to \citet{Alabi2020497}, the centre of galaxy NGC 474 is dominated by slightly lower mass-weighted metallicity [Z/H]\,=\,-0.14$\pm$0.08\,dex (corresponding to 0.72 times of the solar metallicity) while the outside shell has a solar metallicity [Z/H]\,=\,-0.03$\pm$0.09\,dex (corresponding to 0.93 times of the solar metallicity). Thus, it is very likely that SN 2017fgc has a metal-rich progenitor system.
Higher-metallicity WDs could produce relatively more stable, and less radioactive nucleosynthetic products owing to the overabundance of neutrons \citep{Timmes2003590}, which is consistent with the observed evidence that relatively small amount of $^{56}Ni$ are found to be synthesized in the explosion of some HV SNe Ia \citep{Polin2019873,LiWX2021906}.
Moreover, a higher-metallicity companion will produce more-abundant CSM, consistent with recent studies \citep{wangxiaofeng2013Sci,Pan2020895,LiWX2021906}.

\section{Conclusion}

In this paper, we present extensive optical photometric and spectroscopic 
observations of the fast-expanding Type Ia SN 2017fgc. This object can be 
put into the categories of both HVG and HV SNe~Ia according to the classification schemes proposed by \citet{Benetti2005} and \citet{wangxiaofeng2009ApJL}. It has a post-peak decline rate \mb = \dmvalue\,mag and an absolute $B$-band magnitude $M_{\rm max}(B)$ = \mbmag\,mag. Using the radioactive-decay-driven radiation diffusion model \citep{Arnett1982}, we find that SN 2017fgc has a  peak luminosity of $L_{\rm peak}$ = \Lmax\ and a synthesized nickel mass of \Mni = \MniValue.

The light curve and color curve evolution of SN 2017fgc are similar to those of other HV~SNe~Ia such as SN 2002bo and SN 2006X. The relatively bright tails in the $U$ and $B$ light curves observed in SN 2017fgc may be indicators of CSM around the progenitor system. Its spectral evolution is similar to those of SN 2002bo, SN 2006X, and SN 2013gs, with an unusally high \SiII\ velocity near maximum light ($\sim$\,\VSiII) as well as stronger absorptions of \FeII/\MgII\ blends near 4400\,\Angst\ and \FeIII/\FeII/\SiII\ blends near 4900\,\Angst. Moreover, SN 2017fgc and other HV~SNe~Ia are found to have more pronounced secondary maximum peaks in the $Ii$ bands. All of the above features indicate that SN 2017fgc and other HV~SNe~Ia likely have experienced more complete burning, or their progenitors have higher metallicity environments.


Inspection of its birthplace environment indicates that SN 2017fgc was born in a gas bridge with young and metal-rich stellar populations. However, the fact that it was located far away from the center of its host galaxy indicates that its progenitor cannot be metal rich. A possible scenario is that the progenitor of SN 2017fgc could be ejected from the inner part of the companion galaxy during the merger $\sim$2 Gyr ago, or formed as a result of some cold gas remained in the companion disk \citep{Alabi2020497,Fensch2020644}. Detailed study of the host environment of SN 2017fgc is needed. Also, more observations and further modeling are essential to reveal the origin of the strong absorption of \FeII\ and IMEs seen in SN 2017fgc and the nature of the fast-expanding subclass of SNe~Ia. 

\section*{Acknowledgments}
We thank the anonymous referee for his/her suggestive comments that help improve the manuscript a lot. This work is supported by the National Natural Science Foundation of China (NSFC, grants 11873081, U2031209, 12033002, 11633002, 11803076, and 11761141001), the National Program on Key Research and Development Project (grant 2016YFA0400803), and the High Level Talent-Heaven Lake Program of Xinjiang Uygur Autonomous Region of China. Moreover, the Scholar Program of the Beijing Academy of Science and Technology (DZ:BS202002) provides partial support for the work. Staffs of the Lijiang 2.4\,m telescope (LJT), 
the Xinglong 2.16\,m telescope (XLT), and Lick Observatory assisted with the observations. The Chinese Academy of Sciences and the People's Government of Yunnan Province provide support for the LJT, which is corporately 
run and maintained by Yunnan Observatories and the Center for Astronomical Mega-Science (CAS). JuJia Zhang is supported by the National Natural Science Foundation of China (NSFC; grants 11773067 and 11403096), the Youth 
Innovation Promotion Association of the CAS (grant 2018081), and the  Ten 
Thousand Talents Program of Yunnan for Top-notch Young Talents. Support for A.V.F.'s group at U.C. Berkeley was provided by the TABASGO Foundation, the Christopher R. Redlich Fund, and the Miller Institute for Basic Research in Science. Research by D.J.S. is supported by the U.S. National Science Foundation (NSF) grants AST-1821967, 1821987, 1813708, 1813466, 1908972, and by the Heising-Simons Foundation under grant \#2020-1864. J.B., 
D.H., D.A.H., and C.P. were supported by NSF grant AST-1911225. This work 
makes use of data from the Las Cumbres Observatory network.

Some of the observations with the Lick Observatory 1\,m Nickel telescope were conducted by U.C. Berkeley undergraduate students Sanyum Channa, Edward Falcon,  Nachiket Girish, Romain Hardy, Julia Hestenes, Andrew Hoffman, Evelyn Liu, Shaunak Modak, Costas Soler, Kevin Tang, Sameen Yunus, and 
Keto Zhang; we thank them for their excellent work. Lick/KAIT and its ongoing operation were made possible by donations from Sun Microsystems, Inc., the Hewlett-Packard Company, AutoScope Corporation, Lick Observatory, the NSF, the University of California, the Sylvia \& Jim Katzman Foundation, and the TABASGO Foundation. A major upgrade of the Kast spectrograph on the Shane 3\,m telescope at Lick Observatory was made possible through generous gifts from the Heising-Simons Foundation as well as William and Marina Kast. Research at Lick Observatory is partially supported by a generous gift from Google. This research has made use of the services of the ESO Science Archive Facility.

\software{SNooPy2 \citep{Burns2011,Burns2014}, SALT 2.4 \citep{Guy2010523,Betoule2014568}, LOSSPhotPypeline (https://github.com/benstahl92/LOSSPhotPypeline), SN-Spectral Evolution (https://github.com/mwvgroup/SN-Spectral-Evolution), Minim Code \citep{Chatz2013}, IRAF \citep{Tody199352,Tody1986627}, DAOPHOT \citep{daophot}, Photutils \citep{Bradley2020}, lcogtsnpipe \citep{lcogtpipe}, respext(https://github.com/benstahl92/respext), Astropy \citep{Astropy2013},
Matplotlib \citep{Matplotlib2007}, Scipy (https://www.scipy.org/), Numpy (https://numpy.org/)}

\clearpage

\bibliographystyle{aasjournal}
\bibliography{fgc}


\clearpage

\begin{deluxetable*}{lrrrrrrrrrr}
\tablecaption{Photometric Standards in the SN 2017fgc Field 1\tablenotemark{a}}\label{tab:standards}
\tablehead{\colhead{Star} &\colhead{$\alpha$ (J2000)} &\colhead{$\delta$ (J2000)} &\colhead{$U$ (mag)} &\colhead{$B$ (mag)} &\colhead{$V$ (mag)} &\colhead{$R$ (mag)} &\colhead{$I$ (mag)} &\colhead{$g$ (mag)} &\colhead{$r$ (mag)} &\colhead{$i$ (mag)} }
\startdata
1  & 01:20:23.613 & +03:20:23.698 & 16.631(011) & 16.636(006) & 16.087(006) & 15.810(006) & 15.406(008) & 16.305(004) & 15.947(004) & 15.817(005) \\
2  & 01:20:00.124 & +03:19:30.367 & 16.605(011) & 16.792(006) & 16.201(006) & 15.899(006) & 15.465(008) & 16.446(004) & 16.040(004) & 15.884(005) \\
3  & 01:19:58.102 & +03:27:56.473 & 16.538(012) & 16.045(005) & 15.200(005) & 14.758(005) & 14.216(006) & 15.609(004) & 14.914(004) & 14.662(004) \\
4  & 01:20:02.123 & +03:22:04.735 & 16.183(011) & 15.993(006) & 15.311(006) & 14.963(006) & 14.511(006) & 15.615(004) & 15.106(004) & 14.934(004) \\
5  & 01:20:15.741 & +03:27:27.932 & 14.593(007) & 14.396(004) & 13.664(004) & 13.539(013) & 13.904(013) & 14.000(002) & 13.434(004) & 14.888(012) \\
6  & 01:20:24.708 & +03:24:02.095 & 15.114(008) & 15.032(006) & 14.437(006) & 14.207(005) & 14.040(005) & 14.685(004) & 14.275(004) & 14.597(003) \\
7  & 01:19:58.888 & +03:28:12.706 & 17.790(021) & 17.635(007) & 16.934(007) & 16.574(007) & 16.102(010) & 17.250(005) & 16.720(005) & 16.534(005) \\
8  & 01:19:55.296 & +03:21:52.628 & 16.527(011) & 16.420(006) & 15.796(006) & 15.478(006) & 15.045(007) & 16.062(004) & 15.619(004) & 15.465(004) \\
9  & 01:20:16.752 & +03:27:05.422 & 17.123(015) & 17.203(007) & 16.646(007) & 16.369(007) & 15.983(009) & 16.869(005) & 16.503(005) & 16.394(005) \\
10 & 01:19:52.090 & +03:23:27.186 & 15.444(009) & 15.191(006) & 14.508(006) & 14.230(005) & 14.000(005) & 14.812(004) & 14.302(004) & 14.597(003) \\
11 & 01:20:28.530 & +03:24:20.902 & 16.748(012) & 16.181(006) & 15.351(006) & 14.923(006) & 14.420(006) & 15.750(004) & 15.073(004) & 14.858(004) \\
12 & 01:20:04.722 & +03:29:32.716 & 15.742(009) & 15.545(005) & 14.854(005) & 14.497(005) & 14.014(005) & 15.164(004) & 14.644(004) & 14.447(004) \\
13 & 01:19:57.199 & +03:27:11.729 & 17.634(018) & 17.533(007) & 16.874(007) & 16.532(007) & 16.070(009) & 17.163(005) & 16.679(005) & 16.486(005) \\
14 & 01:20:31.904 & +03:20:34.181 & 16.485(011) & 16.426(006) & 15.775(006) & 15.439(006) & 14.967(006) & 16.059(004) & 15.585(004) & 15.393(004) \\
15 & 01:20:22.976 & +03:28:34.205 & 14.711(011) & 14.988(009) & 14.325(009) & 13.898(007) & 13.041(006) & 14.617(006) & 14.129(007) & 13.383(002) \\
16 & 01:20:53.340 & +03:16:40.750 & 13.768(009) & 12.763(001) & 12.248(001) & 11.994(001) & 12.186(009) & 12.444(001) & 12.126(001) & 12.026(001) \\
17 & 01:20:53.291 & +03:26:42.976 & 13.709(010) & 12.387(001) & 11.927(001) & 11.699(001) & 11.956(008) & 12.088(001) & 11.831(001) & 11.733(001) \\
18 & 01:20:44.339 & +03:25:52.392 & 13.672(009) & 12.566(001) & 12.129(001) & 11.913(001) & 12.068(008) & 12.275(001) & 12.045(001) & 11.949(001) \\
19 & 01:19:37.509 & +03:17:06.781 & 14.010(021) & 14.800(014) & 14.010(014) & 13.655(015) & 11.016(011) & 14.383(010) & 13.751(010) & 13.889(011) \\
20 & 01:19:56.437 & +03:29:39.264 & 15.289(008) & 15.176(005) & 14.561(005) & 14.268(005) & 13.901(005) & 14.821(004) & 14.389(004) & 14.363(003) \\
\enddata
\tablenotetext{a}{Standard stars used for calibration of instrumental magnitudes.}
\end{deluxetable*}

\startlongtable
\begin{deluxetable*}{lccccccccccccc}
\tablecolumns{6} \tablewidth{0pc} \tabletypesize{\scriptsize}
\tablecaption{Photometric Observations of SN 2017fgc by Ground-Based Telescopes \label{fgc:gndphot}}
\tablehead{\colhead{MJD} &\colhead{Epoch\tablenotemark{a}} &\colhead{$U$ (mag)} &\colhead{$B$ (mag)} &\colhead{$V$ (mag)} &\colhead{$R$ (mag)} &\colhead{$I$ (mag)} &\colhead{$g$ (mag)} &\colhead{$r$ (mag)} &\colhead{$i$ 
(mag)} &\colhead{$Clear$ (mag)} &\colhead{Telescope}}
\startdata
57947.10 & -12.30 & 15.304(100) & 15.446(043) & 15.381(040) & \nodata  & \nodata  & 15.297(033) & 15.223(033) & 15.742(048) & \nodata  & LCO       
\\
57948.16 & -11.24 & 14.743(076) & 15.010(034) & 14.928(034) & \nodata  & \nodata  & 14.906(027) & 14.857(027) & 15.356(039) & \nodata  & LCO       
\\
57954.72 & -4.68 & 13.635(110) & 13.948(051) & 13.932(051) & \nodata  & \nodata  & 13.893(056) & 13.874(037) & 14.420(063) & \nodata  & LCO       \\
57955.81 & -3.59 & 13.571(052) & 13.875(024) & 13.820(022) & \nodata  & \nodata  & 13.804(019) & 13.868(021) & 14.345(030) & \nodata  & LCO       \\
57956.81 & -2.59 & 13.527(045) & 13.838(019) & 13.673(017) & \nodata  & \nodata  & 13.716(015) & 13.609(015) & 14.316(023) & \nodata  & LCO       \\
57957.94 & -1.46 & 13.545(054) & 13.811(024) & 13.699(022) & \nodata  & \nodata  & 13.711(018) & 13.644(019) & 14.319(030) & \nodata  & LCO       \\
57959.49 & +0.09 & \nodata  & 13.793(034) & 13.596(030) & 13.522(020) & 13.889(031) & \nodata  & \nodata  & \nodata  & \nodata  & Nickel    \\
57959.82 & +0.42 & 13.552(064) & 13.801(028) & 13.655(024) & \nodata  & \nodata  & 13.688(023) & 13.572(022) & 14.337(054) & \nodata  & LCO       \\
57960.49 & +1.09 & \nodata  & 13.819(065) & 13.548(036) & 13.566(048) & 14.038(055) & \nodata  & \nodata  & \nodata  & 13.609(054) & KAIT4     \\
57961.49 & +2.09 & \nodata  & \nodata  & \nodata  & \nodata  & \nodata  & 
\nodata  & \nodata  & \nodata  & 13.589(051) & KAIT4     \\
$\vdots$ & $\vdots$ & $\vdots$ & $\vdots$    & $\vdots$    & $\vdots$    & $\vdots$ & $\vdots$ & $\vdots$ & $\vdots$ & $\vdots$ & $\vdots$     \\
58134.25 & +174.85 & \nodata  & 18.326(128) & 18.276(101) & 18.973(134) & 
19.131(196) & \nodata  & \nodata  & \nodata  & \nodata  & TNT       \\
58136.10 & +176.70 & \nodata  & \nodata  & \nodata  & \nodata  & \nodata  
& \nodata  & \nodata  & \nodata  & 18.341(135) & KAIT4     \\
58137.25 & +177.85 & \nodata  & 18.408(112) & 18.159(086) & 18.790(104) & 
19.342(174) & \nodata  & \nodata  & \nodata  & \nodata  & TNT       \\
58144.11 & +184.71 & \nodata  & 18.659(222) & 18.446(136) & \nodata  & \nodata  & 18.005(119) & 19.192(102) & 19.205(186) & \nodata  & LCO       \\
58146.14 & +186.74 & \nodata  & \nodata  & \nodata  & \nodata  & \nodata  
& \nodata  & \nodata  & \nodata  & 18.629(169) & KAIT4     \\
58150.11 & +190.71 & \nodata  & \nodata  & \nodata  & \nodata  & \nodata  
& \nodata  & \nodata  & \nodata  & 18.427(169) & KAIT4     \\
58152.25 & +192.85 & \nodata  & 18.195(147) & 18.446(108) & 19.218(159) & 
\nodata  & \nodata  & \nodata  & \nodata  & \nodata  & TNT       \\
58154.12 & +194.72 & \nodata  & 18.444(116) & 18.660(182) & 19.690(315) & 
19.112(242) & \nodata  & \nodata  & \nodata  & 18.634(076) & KAIT4     \\
58162.12 & +202.72 & \nodata  & \nodata  & \nodata  & \nodata  & \nodata  
& \nodata  & \nodata  & \nodata  & 18.802(128) & KAIT4     \\
58164.12 & +204.72 & \nodata  & \nodata  & \nodata  & \nodata  & \nodata  
& \nodata  & \nodata  & \nodata  & 18.746(136) & KAIT4     \\
\enddata
\tablenotetext{a}{Relative to the epoch of $B$-band maximum brightness (MJD = 57,959.4). Measurements are calibrated to the AB magnitude system. 
A machine-readable file is available for this table.}
\end{deluxetable*}


\begin{deluxetable*}{lrrrrr}
\tablecaption{Spectroscopic Observations of SN 2017fgc}\label{tab:spectra}
\tabletypesize{\normalsize}
\tablehead{\colhead{MJD}& \colhead{Epoch\tablenotemark{a}} &\colhead{$\lambda_{\rm Start}$} &\colhead{$\lambda_{\rm End}$} &\colhead{Telescope} }
\startdata
57947.7  & -11.7  & 3231 & 9918  & LCO  \\
57952.5  & -6.9   & 3616 & 10,400 & Lick 3\,m \\
57952.6  & -6.8   & 3226 & 9919  & LCO  \\
57959.5  & -0.1   & 3226 & 9919  & LCO  \\
57960.5  & +1.1   & 3622 & 10,400 & Lick 3\,m \\
57961.5  & +2.1   & 3226 & 9919  & LCO  \\
57964.5  & +5.1   & 3620 & 10,400 & Lick 3\,m \\
57966.5  & +7.0   & 3620 & 10,400 & Lick 3\,m \\
57966.5  & +7.1   & 3226 & 9920  & LCO  \\
57972.6  & +13.2  & 3226 & 9919  & LCO  \\
57981.5  & +21.1  & 3231 & 9919  & LCO  \\
57987.5  & +28.4  & 3473 & 9099  & LJT  \\
57988.7  & +29.3  & 3180 & 9919  & LCO  \\
57992.5  & +33.1  & 3622 & 10,400 & Lick 3\,m \\
57997.7  & +38.2  & 3374 & 9919  & LCO  \\
58006.6  & +47.2  & 3972 & 9919  & LCO  \\
58010.5  & +51.1  & 3632 & 10,400 & Lick 3\,m \\
58011.5  & +52.1  & 3634 & 10,400 & Lick 3\,m \\
58013.6  & +54.2  & 3226 & 9919  & LCO  \\
58023.4  & +64.0  & 3277 & 9918  & LCO  \\
58023.5  & +64.1  & 3632 & 10,400 & Lick 3\,m \\
58023.6  & +64.2  & 3276 & 9919  & LCO  \\
58041.5  & +82.1  & 3474 & 9224  & LCO  \\
58045.4  & +86.0  & 3620 & 10,400 & Lick 3\,m \\
58051.4  & +92.0  & 3620 & 10,400 & Lick 3\,m \\
58053.1  & +93.7  & 3701 & 8747  & XLT  \\
58056.4  & +97.0  & 3622 & 10,398 & Lick 3\,m \\
58062.5  & +103.1 & 3276 & 9919  & LCO  \\
58083.1  & +123.7 & 3725 & 8739  & XLT  \\
58091.2  & +131.8 & 3176 & 9918  & LCO  \\
58101.2  & +141.8 & 3227 & 9919  & LCO  \\
58108.0  & +148.6 & 5665 & 6963  & LCO  \\
58109.1  & +149.7 & 2990 & 24,790 & ESO public \\
58116.3  & +156.9 & 3178 & 9902  & LCO  \\
58131.2  & +171.8 & 3226 & 9225  & LCO  \\
58140.3  & +180.9 & 5711 & 7002  & LCO  \\
58343.3  & +383.9 & 2990 & 24,790 & ESO public \\
58348.3  & +388.9 & 2990 & 24,790 & ESO public \\
\enddata
\tablenotetext{a}{Relative to the epoch of $B$-band maximum brightness (MJD = 57,959.4).}
\end{deluxetable*}

\begin{deluxetable*}{lrrrrr}
\tablecaption{DLT40 Photometry of SN 2017fgc in the Clear Band}\label{dlt40phot}
\tabletypesize{\normalsize}
\tablehead{\colhead{MJD}& \colhead{Epoch\tablenotemark{a}} &\colhead{Clear (mag)} &\colhead{Telescope} }
\startdata
57942.3 & -17.1 & 19.183(\nodata) & Prompt5   \\
57943.3 & -16.1 & 17.324(086) & Prompt5   \\
57944.4 & -15.0 & 16.550(040) & Prompt5   \\
57945.3 & -14.1 & 16.096(030) & Prompt5   \\
57946.4 & -13.0 & 15.532(020) & Prompt5   \\
57947.3 & -12.1 & 15.196(027) & Prompt5   \\
57952.3 & -7.1  & 14.130(016) & Prompt5   \\
57953.3 & -6.1  & 14.015(016) & Prompt5   \\
57955.3 & -4.1  & 13.854(016) & Prompt5   \\
57956.3 & -3.1  & 13.779(016) & Prompt5   \\
$\vdots$ & $\vdots$ & $\vdots$ & $\vdots$ \\
58098.1 & 138.7 & 17.728(056) & Prompt5   \\
58099.1 & 139.7 & 17.830(064) & Prompt5   \\
58100.1 & 140.7 & 17.687(060) & Prompt5   \\
58101.1 & 141.7 & 17.814(057) & Prompt5   \\
58102.1 & 142.7 & 17.961(063) & Prompt5   \\
58103.1 & 143.7 & 17.857(060) & Prompt5   \\
58108.1 & 148.7 & 18.137(075) & Prompt5   \\
58110.1 & 150.7 & 17.998(070) & Prompt5   \\
58111.6 & 152.2 & 18.123(067) & Meckering \\
58124.5 & 165.1 & 18.593(079) & Meckering \\
\enddata
\tablenotetext{a}{Relative to the epoch of $B$-band maximum (MJD = 57,959.4). A machine-readable file is available for this table.}
\end{deluxetable*}

\begin{table*}
\centering
\caption{Parameters of SN 2017fgc}
\label{tabpars}
\tabletypesize{\normalsize}
\begin{tabular*}{3.0in}{@{\extracolsep{1in}}ll}
\hline\hline
\multicolumn{1}{c}{Parameter} & \multicolumn{1}{c}{Value} \\\hline
\multicolumn{2}{c}{Photometric}                            \\
$B_{\rm max}$                 & \mB\,mag  \\
$B_{\rm max}-V_{\rm max}$     & \BVmax\,mag \\
$M_{\rm max}(B)$              & \mbmag\,mag \\
$E(B-V)_{\rm host}$           & \ebvalue\,mag   \\
$\Delta m_{15}(B)$            & \dmvalue\,mag \\
$s_{BV}$ 	                   & \sbvalue\     \\
$t_{\rm max}(B)$ 	         & \tbmax\,d \\
$t_0$	                         &	\tzero\,d \\
$\tau_{\rm rise}$	         & \trise\,d \\
$L_{\rm bol}^{\rm max}$	    & \Lmax\ \\
$M_{^{56}\rm Ni}$	          & \MniValue\ \\
\multicolumn{2}{c}{Spectroscopic}                          \\
$v \rm _{0}$(\SiII)	          & \VSiII\ \\
$\dot{v}$(\SiII)		     & \VSiIIDot\ \\
$R$(\SiII)			          & \RSiII\ \\
\hline
\end{tabular*}
\end{table*}

\end{document}